\documentclass[11pt,a4paper]{article}
\usepackage{amsmath}
\usepackage{amsfonts}
\usepackage{color}
\usepackage{graphicx}
\usepackage{dcolumn}
\usepackage{bm}
\usepackage[
bookmarks,bookmarksnumbered,colorlinks=true,anchorcolor=blue,
linkcolor=blue,urlcolor=blue,citecolor=blue,
breaklinks=true]{hyperref}
\usepackage{geometry}
\usepackage{authblk}
\usepackage{amssymb}
\usepackage[utf8]{inputenc}

\def\k{{\kappa}}
\def\l{{\lambda}}

\def\d{{\delta}}
\def\D{{\Delta}}
\def\o{{\omega}}
\def\O{{\Omega}}
\def\e{{\epsilon}}

\def\a{{\alpha}}
\def\b{{\beta}}
\def\c{{\chi}}

\def\g{{\gamma}}

\def\m{{\mu}}
\def\n{{\nu}}

\def\s{{\sigma}}

\def\x{{\xi}}

\newcommand{\lag}{\langle}
\newcommand{\rag}{\rangle}

\newcommand{\pd}{{\partial}}

\date{\today}

\begin{document}
\title{\bf Magneto-vortical effect in strongly coupled plasma}

\author[1]{Yanyan Bu \thanks{yybu@hit.edu.cn}}
\author[2]{Shu Lin \thanks{linshu8@mail.sysu.edu.cn}}
\affil[1]{\it School of Physics, Harbin Institute of Technology, Harbin 150001, China} 
\affil[2]{\it School of Physics and Astronomy, Sun Yat-Sen University, Zhuhai 519082, China }

\maketitle

\begin{abstract}
Based on a holographic model incorporating both chiral anomaly and gravitational anomaly, we study the effect of magneto-vortical coupling on transport properties of a strongly coupled plasma. The focus of present work is on the generation of a vector charge density and an axial current, as response to vorticity in a magnetized plasma. The transport coefficients parameterising the vector charge density and axial current are calculated both analytically (in the weak magnetic field limit) and also numerically (for general values of the magnetic field). We find the generation of vector charge receives both non-anomalous and anomalous contributions, with the non-anomalous contribution dominating in the limit of strong magnetic field and the anomalous contribution sensitive to both chiral anomaly and gravitational anomaly. On the contrary, we find the axial current is induced entirely due to the gravitational anomaly, thus we interpret the axial current generation as chiral vortical effect. The corresponding chiral vortical conductivity is found to be suppressed by the magnetic field. By Onsager relation, these transport coefficients are responsible for the generation of a thermal current due to a transverse electric field or a transverse axial magnetic field, which we call thermal Hall effect and thermal axial magnetic effect, respectively.
\end{abstract}

\newpage


\allowdisplaybreaks

\flushbottom

\section{Introduction}

The effect of magnetic field and vorticity to QCD matter has attracted much attention over the past few years. At very high temperature, when quarks become asymptotically free, a charged-neutral QCD matter can be either magnetized in magnetic field or polarized in vorticity field. Close to the chiral phase transition, when the interaction among quarks becomes strong, more interesting phenomena such as inverse magnetic catalysis \cite{Bali:2012zg,Bali:2011qj} and vector meson condensation \cite{Chernodub:2011mc,Chernodub:2010qx} can emerge. Similarly, vorticity field may suppress the chiral condensation \cite{Jiang:2016wvv,Ebihara:2016fwa}.
%

When the QCD matter carries net vector charge or axial charge densities, the chiral anomaly and gravitational anomaly can induce a variety of anomalous transport phenomena such as the chiral magnetic effect (CME) \cite{Vilenkin:1980fu,Fukushima:2008xe,Fukushima:2010vw}, the chiral vortical effect (CVE) \cite{Erdmenger:2008rm,Banerjee:2008th,Son:2009tf} and the chiral separation effect (CSE) \cite{Metlitski:2005pr,Son:2004tq}, etc.

Recently, the interplay of a strong magnetic field and a vorticity is found to lead to new transport phenomenon such as dynamical generation of a vector charge \cite{Hattori:2016njk}, see also \cite{Liu:2017spl,Chen:2015hfc,Cao:2019ctl,Chen:2019tcp}. Under the lowest Landau level (LLL) approximation, Hattori and Yin found the generation of a vector charge from spin-vorticity coupling as \cite{Hattori:2016njk}\footnote{see also \cite{Ebihara:2016fwa} for possible contribution from orbital angular momentum and vorticity coupling.}
\begin{align} \label{Jt_LLL}
 J^t= q_f \frac{C_A}{2} (\vec B\cdot \vec \O),
\end{align}
with $C_A=\frac{1}{2\pi^2}$ the chiral anomaly coefficient.
In fact, such a contribution should be viewed as a large $B$, free limit of a QED plasma. More generally, one would expect from the viewpoint of polarisable matter \cite{Kovtun:2016lfw} that:
\begin{align} \label{Jt_expect}
  J^t= \xi(B,T) (\vec B\cdot \vec \O), 
\end{align}
Moreover, if we could associate an effective chemical potential for the generated vector charge, 
this can further give rise to the generation of an axial current by the chiral anomaly and gravitational anomaly. Note that the vector charge susceptibility is $\c=C_A|q_f|B$ in the LLL approximation, and thus \eqref{Jt_LLL} corresponds to the effective chemical potential $\m_{\rm eff}=\frac{{\rm sgn}(q_f)}{2}{\vec\Omega}\cdot{\hat B}$ with $\hat B= \vec B/|\vec B|$. The vector charge imbalance would result in an axial current through the chiral separation effect \cite{Metlitski:2005pr,Son:2004tq}
\begin{align}\label{J5_LLL}
  \vec J_5= |q_f| \frac{C_A}{2} (\vec B\cdot \vec \O) \hat{B}.
\end{align}
Again, this is a large $B$, free limit of a QED plasma. More generally, one would expect an extra contribution from the gravitational anomaly, which always induces a temperature-dependent contribution to the axial current even in the absence of the chiral imbalance \cite{Son:2009tf,Neiman:2010zi,Landsteiner:2011cp}. Therefore, we expect a more general axial current
\begin{align} \label{J5_expect}
  \vec J_5 = \sigma(B,T) \vec \O.
\end{align}

It is worth noting that the physical picture behind \eqref{Jt_LLL} and \eqref{J5_LLL} is the spectral flow: a shift in background vector gauge field leads to opposite energy shift for right- and left-handed fermions, generating net {\it axial} charge. In order for the spectral flow picture to generate vector charge, we would need an axial gauge field, whose coupling to right- and left-handed fermions differ in sign, thus leading to the same energy shift for them. In the analysis of Hattori and Yin \cite{Hattori:2016njk}, the role of an axial gauge field is played by a vorticity. Indeed, in free theory, we have ${\vec S}={\vec J}_5$ so that we can identify ${\vec A}_5$ with ${\vec \O}$ by comparing the coupling ${\vec\O}\cdot{\vec S}$ with ${\vec J}_5\cdot{\vec A}_5$. However, in an interacting theory, the ``equivalence'' of ${\vec A}_5$ with ${\vec\O}$ is far from obvious.
First of all, even in a free theory the presence of an axial gauge field as a source poses an ambiguity in the definition of currents: consistent current and covariant current could differ by terms proportional to the axial gauge field \cite{Landsteiner:2016led}. Similar ambiguity does not exist in the case with the vorticity as a source. Secondly, in an interacting theory, the vorticity couples to the angular momentum as a whole. The separation of the spin from the total angular momentum is often ambiguous. Therefore, it is desirable to go away from the free theory limit to test the robustness of the mechanism. In this paper, we go to the opposite limit, where the theory is strongly coupled. Specifically, we will study the response of a strongly coupled magnetized plasma to the vorticity field by a holographic model.

The rest of the paper is organized as follows: In Section \ref{holo_setup}, we present the setup of the holographic model. In Section \ref{fluc_analysis}, we turn on a metric perturbation as a proxy for the vorticity in the magnetized plasma. We will study the response of the vector charge density and axial current to the vorticity. In Section \ref{caseI_study}, we will present both analytic results in small $B$ regime and numerical results for general $B$. 
In Section \ref{hydro_analysis}, we use the Onsager relation to obtain thermal Hall effect and thermal axial magnetic effect. We conclude and discuss implications of our results in Section \ref{conclusion}. Details of the computations are collected in appendices \ref{details_background}, \ref{horizon_matching} and \ref{app_suscept}.

\section{Holographic setup: magnetic brane in $AdS_5$} \label{holo_setup}

\subsection{Gravity Action and Dictionary} \label{action_dic}
We extend the holographic model initially considered in \cite{1006.2400,1107.0368} by including both vector and axial gauge fields. The full action is
\begin{equation} \label{bulk_action}
\begin{split}
S&=\frac{1}{2\kappa^2}\int d^5x \sqrt{-g}\left\{R[g]+12-\frac{1}{4}(F^V)^2- \frac{1}{4} (F^a)^2+  \epsilon^{MNPQR} A_M \right.\\
&\left.\times\left[\frac{1}{3}\alpha (F^a)_{NP} (F^a)_{QR} + \alpha(F^V)_{NP}(F^V)_{QR}+ \lambda R^Y_{~XNP} R^X_{~~YQR} \right] \right\}\\
&+\frac{1}{\kappa^2}\int d^4x \sqrt{-\gamma} K[\gamma] + \underline{S_{\rm{CSK}}}+ S_{\rm c.t.},
\end{split}
\end{equation}
where $F^V=dV$ and $F^a=dA$. The last line of \eqref{bulk_action} correspond to boundary terms defined on the hypersurface $\Sigma$ of constant $r$.
The notation $\gamma$ denotes the determinant of the induced metric $\gamma_{\mu\nu}$ on $\Sigma$:
\begin{equation}
ds^2|_{\Sigma}=g_{MN}dx^M dx^N|_{\Sigma}=\gamma_{\mu\nu} dx^\mu dx^\nu.
\end{equation}
We also need the out-pointing unit normal vector of the surface $\Sigma$:
\begin{equation}
n_M= \frac{\partial_M r}{\sqrt{g^{AB} \partial_A r \partial_B r}}.
\end{equation}
Moreover, $K=\gamma^{\mu\nu} K_{\mu\nu}$ whereas $K_{\mu\nu}$ is the extrinsic curvature tensor
\begin{equation}
K_{\mu\nu}=  \frac{1}{2} \mathfrak{L}_n \gamma_{\mu\nu}=\frac{1}{2}\left( n^M \partial_M \gamma_{\mu\nu} +\gamma_{\mu N} \partial_\nu n^N+ \gamma_{\nu N} \partial_\mu n^N\right).
\end{equation}
The Levi-Civita tensor is $\epsilon^{MNPQR}= \epsilon(MNPQR)/\sqrt{-g}$ whereas $\epsilon(MNPQR)$ is the Levi-Civita symbol under the convention $\epsilon(rtxyz)=+1$. The purely gauge Chern-Simons action ($\alpha$-terms) mimics the chiral anomaly while the mixed gauge-gravitational Chern-Simons term ($\lambda$-term) is to model the gravitational anomaly of the boundary field theory.

As explained in \cite{1107.0368}, in order to get a correct form of gravitational anomaly (i.e. guarantee the gauge variation of the bulk action to be a total derivative), one needs to add the term
\begin{equation} \label{CSK}
S_{\rm{CSK}}=-\frac{4}{\kappa^2}\lambda \int d^4x \sqrt{-\gamma} n_M \epsilon^{MNPQR} A_N K_{PL} \tilde{\nabla}_Q K_R^L,
\end{equation}
where $\tilde{\nabla}$ is compatible with the induced metric $\gamma_{AB}$.
The counter-term action is
\begin{equation} \label{counter}
S_{\rm c.t.}=-\frac{1}{2\kappa^2} \int d^4x \sqrt{-\gamma}\left(6+ \frac{1}{2} R[\gamma] -\mathcal{C}_t\right),
\end{equation}
where $\mathcal{C}_t$ cancels the logarithmic divergences \cite{deHaro:2000vlm,Sahoo:2010sp}
\begin{equation}
\begin{split}
\mathcal{C}_t&= \frac{1}{4} \log r \left[\left(F^V\right)_{\mu\nu} \left(F^V\right)^{\mu\nu} + \left(F^a\right)_{\mu\nu} \left(F^a\right)^{\mu\nu} \right]\\
&+ \log\frac{1}{r^2} \left(\frac{1}{8} R^{\mu\nu}[\gamma]R_{\mu\nu}[\gamma] -\frac{1}{24} R^2[\gamma]\right).
\end{split}
\end{equation}
Note that $\mathcal{C}_t$ non-vanishes only when nontrivial sources (either external gauge fields or non-flat boundary metric) are turned on for the boundary theory. In addition, in $\mathcal{C}_t$ we employ the minimal subtraction scheme so that it will not generate {\it finite} contribution to the boundary currents and stress tensor.

According to the holographic dictionary, expectation values of the stress tensor and currents of the boundary theory are defined as
\begin{equation}
T_{\mu\nu}\equiv  \lim_{r\to\infty} \frac{-2r^2}{\sqrt{-\gamma}} \frac{\delta S}{\delta \gamma^{\mu\nu}},\qquad \qquad J^\mu\equiv \lim_{r\to\infty} \frac{\delta S}{\delta V_\mu} ,\qquad \qquad J^\mu_5\equiv \lim_{r\to\infty} \frac{\delta S}{\delta A_\mu},
\end{equation}
Explicitly, the vector current is (from here one, we set $2\k^2=1$ for convenience):
\begin{equation} \label{Jmu_cov}
J^\mu= \lim_{r\to\infty} \sqrt{-\gamma} \left\{n_M\left(F^V\right)^{\mu M}+ 4\alpha n_M \epsilon^{M\mu N QR} A_N \left(F^V\right)_{QR} -\tilde{\nabla}_\nu \left(F^V\right)^{\nu\mu} \log r \right\}.
\end{equation}
However, the axial current and stress tensor are somehow subtle/complicated:
\begin{equation} \label{J5mu_cov}
J_5^\mu =\lim_{r\to\infty} \sqrt{-\gamma} \left\{n_M\left(F^a\right)^{\mu M}+ \frac{4}{3}\alpha n_M \epsilon^{M\mu N QR} A_N \left(F^a\right)_{QR} -\tilde{\nabla}_\nu \left(F^a\right)^{\nu\mu} \log r +J_{\rm CSK}^\mu\right\},
\end{equation}
\begin{equation}
T_{\mu\nu}=-2\lim_{r\to\infty}r^2\left(K_{\mu\nu}-K\gamma_{\mu\nu} +3\gamma_{\mu\nu} - \frac{1}{2} \mathcal{G}_{\mu\nu}[\gamma] - T_{\mu\nu}^{\rm Gra}\right) + T_{\mu\nu}^{\mathcal C},
\end{equation}
where $T_{\mu\nu}^{\mathcal C}$ arises from the functional derivative of $\mathcal{C}_t$,  $J_{\rm CSK}^\mu$ is due to the added action $S_{\rm CSK}$, and $T_{\mu\nu}^{\rm Gra}$ comes from the gravitational Chern-Simons term.
The expressions for all of them are \cite{1107.0368,1304.5529}:
\begin{align}
& J_{\rm CSK}^\mu=-8\lambda n_M \epsilon^{M\mu PQR}K_{PL} \tilde \nabla_Q K^L_R, \nonumber \\
& T_{\mu\nu}^{\mathcal C}= T_{\mu\nu}^{\mathcal C_1}+ \lim_{r\to \infty} \frac{1}{4} r^6 \log r \left[\gamma_{\mu\nu} (F^V)_{\alpha\beta} (F^V)^{\alpha\beta} - 4 (F^V)_{\mu\alpha} (F^V)_\nu^{~\alpha}\right]+ (V\rightarrow a), \nonumber \\
& T^{\mu\nu}_{\rm Gra}= 4\lambda \epsilon^{(\mu\alpha\beta\rho}\left[\frac{1}{2} (F^a)_{\alpha\beta}R^{\nu)}_{~\rho}[\gamma] +\tilde \nabla_\delta \left(A_\alpha R^{\delta \nu)}_{~~\beta\rho}[\gamma] \right) \right],
\end{align}
where $T_{\mu\nu}^{\mathcal C_1}$ vanishes for a flat boundary. $T^{\mu\nu}_{\rm Gra}$ was first derived in \cite{1304.5529} based on the ADM decomposition approach.
Above, we stick to the consistent current formalism. Indeed, in the absence of a background for the axial gauge field, there will be no difference between the consistent current and covariant current \cite{Landsteiner:2016led}.
The authors of \cite{1006.2400} presented thorough analysis for the holographic renormalisation of the model, but did not get the term $S_{\rm CSK}$. Additionally, the authors of \cite{1006.2400} addressed that the gravitational Chern-Simons term will make contribution to the boundary stress tensor. See also \cite{1304.5529,1706.05294} for more recently updated formulas for stress tensor and axial current of the boundary theory. The holographic model does correctly describe the chiral/gravitational anomalies for the boundary field theory \cite{1107.0368}:
\begin{align}
\hat\nabla_\mu J^\mu=0,\qquad \hat \nabla_\mu J_5^\mu= 8\alpha \hat\epsilon^{\alpha\beta\rho\delta} \hat F_{\alpha \beta} \hat F_{\rho \delta}+ \lambda \hat\epsilon^{\alpha\beta\rho\delta} \hat R^\tau_{~\kappa \alpha \beta} \hat R^\kappa_{~ \tau \rho\delta},
\end{align}
where a hat is to remind that the corresponding quantity is defined on the boundary.

Under the variation
\begin{equation}
g_{MN}\to g_{MN}+\delta g_{MN},\qquad V_M\to V_M+\delta V_M,\qquad A_M\to A_M+ \delta A_M,
\end{equation}
one obtains the Einstein equation
\begin{equation}
0=E_{MN}\equiv R_{MN}-\frac{1}{2}R g_{MN}-6 g_{MN}- T_{MN}^{\rm bulk},
\end{equation}
and anomalous Maxwell equations
\begin{equation}
\begin{split}
0=EV^M \equiv \nabla_N\left(F^V\right)^{NM}+ 2\alpha \epsilon^{MNPQR} \left(F^a \right)_{NP} \left(F^V\right)_{QR},
\end{split}
\end{equation}
\begin{equation}
\begin{split}
0=EA^M\equiv & \nabla_N \left(F^a\right)^{NM} + \alpha \epsilon^{MNPQR} \left[\left(F^V\right)_{NP} \left(F^V\right)_{QR} +\left(F^a\right)_{NP} \left(F^a\right)_{QR}\right] \\
&+ \lambda\epsilon^{MNPQR}  R^Y_{~XNP} R^X_{~~YQR}.
\end{split}
\end{equation}
The bulk stress tensor $T_{MN}^{\rm bulk}$ could be split into two parts
\begin{equation}
T_{MN}^{\rm bulk}= T_{MN}^{\rm Maxwell}- \nabla_X \Theta^X_{MN},
\end{equation}
where
\begin{equation}
T_{MN}^{\rm Maxwell}=\frac{1}{2} \left(F^V\right)_{RM} \left(F^V\right)^R_{~N} -\frac{1}{8} g_{MN} \left(F^V\right)^2 + \frac{1}{2} \left(F^a\right)_{RM} \left(F^a\right)^R_{~N} -\frac{1}{8} g_{MN} \left(F^a\right)^2
\end{equation}
\begin{equation}
\Theta^X_{MN}=\lambda \epsilon^{QRSTU}\left(g_{QM} R^X_{~NRS}+ g_{QN} R^X_{~MRS}\right) \left(F^a\right)_{TU}.
\end{equation}
Alternatively, the Einstein equation could be rewritten as
\begin{align}
0=E_{MN}=R_{MN}+4g_{MN}-\tilde{T}^{\rm bulk}_{MN}
\end{align}
where
\begin{align}
\tilde{T}^{\rm bulk}_{MN}&=\frac{1}{2}(F^V)_{RM}(F^V)^R_{~N} -\frac{1}{12}g_{MN}(F^V)^2 + \frac{1}{2}(F^a)_{RM} (F^a)^R_{~N}-\frac{1}{12}g_{MN}(F^a)^2 \nonumber\\
& - \nabla_X \Theta^X_{MN}+ \frac{1}{3}g_{MN}g^{AB}\nabla_X \Theta^X_{AB}.
\end{align}

\subsection{Neutral Magnetic Brane Background} \label{magnetic_brane}
To proceed, we consider the background solution of the holographic model \eqref{bulk_action}. For simplicity, we focus on a neutral magnetized plasma. To this end, we turn on a constant magnetic field along the $z$-direction,
\begin{equation} \label{V background}
V=Bx dy \Rightarrow \vec{B}=B \hat{z},
\end{equation}
which obviously breaks the $SO(3)$ rotational symmetry to $SO(2)_\perp$ on the $xy$-plane. As a result, the background metric takes the form
%
\begin{equation} \label{bg metric}
ds^2=2drdt-f(r)dt^2+ e^{2W_T(r)}(dx^2+dy^2)+ e^{2 W_L(r)} dz^2.
\end{equation}
Note that in writing down (\ref{bg metric}), the ingoing Eddington-Finkelstein coordinate has been employed in order to avoid coordinate singularity.
The background metric (\ref{bg metric}) has an event horizon at $r=r_h$ so that
\begin{equation}
f(r\simeq r_h)=0+f^\prime(r_h)(r-r_h)+\cdots,
\end{equation}
while $W_T,W_L$ are regular at $r=r_h$. The Hawking temperature, identified as the temperature of the dual gauge theory, is
\begin{equation}
T= \frac{\partial_r(f(r))}{4\pi}\Bigg|_{r=r_h}.
\end{equation}
Generically, both $W_T(r)$ and $W_L(r)$ will depend on $r_h$ nontrivially.

It is a simple exercise to check that, given above ansatz \eqref{V background} and \eqref{bg metric}, both the gauge and gravitational Chern-Simons terms do not affect the bulk equations of motion. Therefore, the background geometry is simply the ``magnetic brane'' solution initially studied in \cite{0908.3875}.

The ordinary differential equations (ODEs) for the metric functions in (\ref{bg metric}) are
\begin{align}
&E_{rr}=0:\quad &0=W_L^{\prime\;2}+2W_T^{\prime\;2}+W_L^{\prime\prime}+2W_T^{\prime \prime}, \label{Wtl}\\
&E_{rt}=0=E_{tt}:\quad &24+B^2e^{-4W_T(r)}=3f^\prime (W_L^\prime+2W_T^\prime)+ 3f^{\prime\prime}, \label{f}\\
&E_{xx}=0=E_{yy}:\quad &12-B^2e^{-4W_T(r)}=3f^\prime W_T^\prime +3f(W_L^\prime W_T^\prime +2W_T^{\prime\;2}+ W_T^{\prime\prime}), \label{Wt}\\
&E_{zz}=0:\quad &24+B^2e^{-4W_T(r)}=6f^\prime W_L^\prime +6f(W_L^{\prime\;2}+ 2W_L^\prime W_T^{\prime} + W_L^{\prime\prime}), \label{Wl}
\end{align}
where the prime denotes a derivative with respect to $r$. The equations \eqref{Wt} and \eqref{Wl} look different from those of \cite{0908.3875}. However, suitable combinations of above equations give rise to the results of \cite{0908.3875}:
\begin{align}
2\times (\ref{Wt})-(\ref{Wl})& \Rightarrow 2f(W_T^{\prime\prime}-W_L^{\prime\prime}) + 2\left[f^\prime +f(W_L^\prime+2 W_T^\prime)\right](W_T^\prime- W_L^{\prime})=-B^2 e^{-4W_T(r)}, \nonumber\\
4\times(\ref{Wt})+(\ref{Wl}) &\Rightarrow 2f^\prime(2W_T^\prime+ W_L^\prime) +4f W_T^\prime (W_T^\prime+ 2 W_L^\prime)=24-B^2 e^{-4W_T(r)}.
\end{align}
Obviously, not all the equations in \eqref{Wtl}-\eqref{Wl} are independent: we will take \eqref{Wtl} as the constraint and solve all the rest to determine the metric functions $f(r),W_T(r),W_L(r)$.

In order to fully determine $f(r),W_T(r),W_L(r)$, we have to impose two boundary conditions for each of them. For $f(r)$, we impose
\begin{align}
f(r=r_h)=0;\qquad \qquad \underline{f(r\to \infty) \to r^2}.
\end{align}
However, it is found that the second boundary condition (underlined above) is automatically satisfied by the bulk EOMs. This demands one to impose another condition for $f(r)$, which is explained in appendix \ref{details_background}.
For $W_T(r)$ and $W_L(r)$, the boundary conditions are
\begin{align}
&W_T(r),\,\, W_L(r) \to \log r, \qquad \qquad \quad \,\,\,\, {\rm as} \quad r\to \infty, \label{AdS_requirement}\\
&12-B^2 e^{-4 W_T(r)}= 3 f^\prime W_T^\prime, \qquad \qquad {\rm at}\quad r=r_h, \label{horizon_Wt}\\
&24+ B^2 e^{-4 W_T(r)}= 6 f^\prime W_L^\prime, \qquad \qquad {\rm at}\quad r=r_h, \label{horizon_Wl}
\end{align}
where the last two equations are read off from \eqref{Wt} and \eqref{Wl} by requiring regularity of $W_T(r), W_L(r)$ at the horizon $r=r_h$.

We solve the bulk EOMs \eqref{Wtl}-\eqref{Wl} analytically when the magnetic field is weak (i.e., $B/T^2\ll 1$) and numerically when $B$ is general. The calculational details as well as the main results are deferred to appendix \ref{details_background}.

\section{Fluctuation in the bulk theory: general consideration} \label{fluc_analysis}

\subsection{Bulk Perturbations} \label{fluc_eom}
In this section, we study the linear response of the magnetized plasma to a fluid vorticity.
A weak fluid vorticity $\vec \O$ would be mimicked by a gravito-magnetic field \cite{1107.0368,1706.05294}. More precisely, one perturbs the boundary Minkowski spacetime (where the fluid flows) as
\begin{align}
ds^2_{\rm M}= -dt^2 + d{\vec x}^2 + 2 h_{ti}(t,\vec x)dt dx^i, \qquad {\rm with}\qquad h_{ti}(t,\vec x)=u_i(t,\vec x).
\end{align}
Then, the vorticity is generated at linear order in $h_{ti}$ as
\begin{align} \label{vorticity}
  \O^i=\frac{1}{2}\e^{ijk}\nabla_ju_k=\frac{1}{2}\e^{ijk}\pd_jh_{tk},
\end{align}
with the unperturbed fluid velocity $u^\m=(1,0,0,0)$. Thus, the curl of $h_{ti}$ could be thought of as a fluid vorticity. 
We take
\begin{align}
h_{ti}(\vec x)=e^{iqx} h_{ty}(q) \delta_{yi},
\end{align}
which gives rise to a stationary vorticity along the $z$-direction, i.e. parallel to the magnetic field. Then, we can obtain the Kubo formulas for the transport coefficients defined in \eqref{Jt_expect} and \eqref{J5_expect} as
\begin{align} \label{kubo}
\xi= \frac{2}{B}\lim_{q\to 0}\frac{\langle J^t T^{ty}\rangle}{iq}, \qquad \qquad
\sigma= 2 \lim_{q\to 0} \frac{\langle J^z_5 T^{ty}\rangle}{iq}.
\end{align}
These will be used in holographic calculations.

To turn on a gravito-magnetic field in the bulk, it is convenient to use the Poincare coordinate system so that the bulk metric takes a diagonal form
\begin{equation} \label{metric_diagonal1}
ds^2=-f(r)dt^2 + \frac{dr^2}{f(r)}+e^{2W_T(r)}\left(dx^2+dy^2\right)+ e^{2W_L(r)} dz^2,
\end{equation}
where we still denote the time of the bulk theory by $t$.
On top of the background \eqref{metric_diagonal1} and \eqref{V background}, it is consistent to turn on the following fluctuation modes
\begin{align} \label{fluc_kubo1}
&\delta(ds^2)= 2 e^{W_T(r)}\left[\delta g_{ty}(r,t,x)dtdy + \delta g_{xy}(r,t,x) dxdy \right], \nonumber\\
&\delta V= \delta V_t(r,t,x) dt + \delta V_x(r,t,x)dx , \qquad \delta A= \delta A_z(r,t,x) dz,
\end{align}
while setting all the rest corrections to zero. Here, we have assumed $(t,x)$-dependent fluctuations for the reason to be discussed in the next subsection and consider a plane wave ansatz:
%
%
%
\begin{align}
&\delta g_{ty}(r,t,x)\sim e^{-i\omega t+iq x} \delta g_{ty}(r), \qquad \qquad  \delta g_{xy}(r,t,x) \sim e^{-i\omega t+iq x} \delta g_{xy}(r), \nonumber \\
&\delta V_t(r,t,x) \sim e^{-i\omega t+i qx} \delta V_t(r), \qquad \qquad \quad \delta V_x(r,t,x) \sim e^{-i\omega t+i qx} \delta V_x(r), \nonumber \\
&\delta A_z(r,t,x) \sim e^{-i\omega t+i qx} \delta A_z(r).
\end{align}
In what follows we record the bulk equations of motion for the fluctuation modes.
First, we consider the constraint equations. The constraint $E_{ry}=0$ is
\begin{align} \label{cons_Ery}
0=& \lambda \omega q e^{-W_L} \delta A_z\left(-\frac{2B^2 e^{-4W_T}}{f} + \frac{2f^\prime W_L^\prime}{f} + \frac{4f^\prime W_T^\prime}{f} -4W_L^\prime W_T^\prime - 8 W_T^{\prime 2} \right) \nonumber \\
&+\frac{1}{2}iq \partial_r\delta g_{xy}+ \frac{i\omega e^{2W_T}}{2f} \partial_r\delta g_{ty} + \frac{1}{2} Be^{-2W_T} \partial_r \delta V_x.
\end{align}
The constraint $EV^r=0$ gives
\begin{align} \label{cons_EVr}
0=8\omega \alpha B \delta A_z + q e^{W_L} f \partial_r \delta V_x + \omega e^{W_L +2 W_T} \partial_r \delta V_t.
\end{align}
Next, we turn to the dynamical components of the bulk EOMs. The Einstein equation $E_{ty}=0$ reads:
\begin{align} \label{del_gty2}
0= &\partial_r\left(e^{W_L+4W_T} \partial_r \delta g_{ty}\right)- \frac{B^2 e^{W_L}} {f(r)} \delta g_{ty}- \frac{q^2 e^{W_L+2W_T}}{f(r)} \delta g_{ty} -\frac{\omega q e^{W_L+2W_T}}{f} \delta g_{xy} \nonumber\\
&+ \frac{i\omega B e^{W_L}}{f} \delta V_x
+ \frac{i q B e^{W_L}}{f(r)} \delta V_t + iq \lambda e^{2W_T} \partial_r \delta A_z \left(4B^2 e^{-4W_T} -4 f^\prime W_L^\prime -8 f^\prime W_T^\prime\right. \nonumber \\
&\left. + 8 f W_T^\prime W_L^\prime +16 f W_T^{\prime 2}\right) + iq \lambda e^{2W_T} \delta A_z \left(2B^2 e^{-4W_T} f^{-1} f^\prime -48 f^{-1}f^\prime- 4B^2 e^{-4W_T} W_L^\prime \right. \nonumber\\
&\left.+ 4f^{-1} f^{\prime2}W_L^\prime + 8f^\prime W_L^{\prime2} -20 B^2 e^{-4W_T} W_T^\prime +96 W_T^\prime + 8 f^{-1} f^{\prime2} W_T^\prime +16 f^\prime W_T^\prime W_L^\prime\right. \nonumber \\
&\left.  -16 f W_T^\prime W_L^{\prime2}- 48 f W_L^\prime W_T^{\prime2} -32 f W_T^{\prime3} \right).
\end{align}
The Einstein equation $E_{xy}=0$ is
\begin{align} \label{del_gxy2}
0= \partial_r\left(e^{W_L+2W_T} f\partial_r \delta g_{xy}\right) + \frac{\omega^2 e^{W_L+ 2W_T}}{f} \delta g_{xy} + \frac{\omega q e^{W_L+2W_T}}{f} \delta g_{ty}.
\end{align}
The Maxwell equation $EV^t=0$ is
\begin{align} \label{del_Vt2}
0=& \partial_r\left(e^{W_L+2W_T}\partial_r \delta V_t\right) + 8\alpha B \partial_r \delta A_z - \frac{\omega q e^{W_L}}{f} \delta V_x- \frac{q^2 e^{W_L}}{f}\delta V_t -\frac{iq B e^{W_L}}{f}\delta g_{ty}
\end{align}
The Maxwell equation $EV^x=0$ is
\begin{align} \label{del_Vx2}
0= \partial_r\left(e^{W_L}f \partial_r \delta V_x\right) + \frac{\omega^2 e^{W_L}}{f}\delta V_x + \frac{\omega q e^{W_L}}{f} \delta V_t + \frac{i\omega B e^{W_L}}{f} \delta g_{ty}.
\end{align}
Finally, $EA^z=0$ yields
\begin{align} \label{del_Az2}
0= & \partial_r\left(e^{2W_T-W_L}f \partial_r \delta A_z\right) + \frac{\omega^2 e^{2W_T-W_L}}{f} \delta A_z - q^2 e^{-W_L}\delta A_z + iq \lambda e^{2W_T} \left(-4 B^2 e^{-4W_T} \right. \nonumber\\
&\left.+ 4 f^\prime W_L^\prime + 8 f^\prime W_T^\prime - 8f W_T^\prime W_L^\prime -16 f W_T^{\prime2}\right)\partial_r \delta g_{ty}+ 8\alpha B \partial_r \delta V_t.
\end{align}

\subsection{Adiabatic Limit and Boundary Conditions} \label{adia_boundary_condition}

From the Kubo formulas \eqref{kubo}, it seems as if we could set $\o=0$ from the beginning. However the boundary condition at the horizon cannot be uniquely determined in this case. This ambiguity is related to the ambiguity in the Kubo formula itself. It can be evaluated in any equilibrium state, charged one or neutral one. For our purpose, it should be evaluated in the unperturbed neutral plasma state.
The boundary condition to use should correspond to the neutral state.
In practice, we specify the state as follows: the state is realized by turning on the vorticity field adiabatically to the original neutral magnetized plasma.

We will seek solutions to \eqref{cons_Ery} through \eqref{del_Az2} in the adiabatic limit $\o\to0$. To this end, we expand the bulk perturbations in powers of $\o$:
\begin{align}\label{omega_exp}
  X=X^{(0)}+\o X^{(1)}+\cdots,
\end{align}
with $X=\d g_{xy},\;\d V_x,\;\d g_{ty},\;\d V_t$ and $\d A_z$. In fact, we only need the leading order solution $X^{(0)}$, for which we suppress the superscript $(0)$. The fields decouple into two sets $\{\d g_{xy},\;\d V_x \}$ and $\{\d g_{ty},\;\d V_t,\;\d A_z\}$. The set $\{\d g_{ty},\;\d V_t,\;\d A_z\}$ satisfies the following equations
\begin{align}
  &\partial_r\left(e^{W_L+4W_T} \partial_r \delta g_{ty}\right)- \frac{B^2 e^{W_L}} {f(r)} \delta g_{ty}- \frac{q^2 e^{W_L+2W_T}}{f(r)} \delta g_{ty}
+ \frac{i q B e^{W_L}}{f(r)} \delta V_t \nonumber \\
&+ iq \lambda e^{2W_T} \partial_r \delta A_z \left(4B^2 e^{-4W_T} -4 f^\prime W_L^\prime -8 f^\prime W_T^\prime + 8 f W_T^\prime W_L^\prime +16 f W_T^{\prime 2}\right) \nonumber \\
&+ iq \lambda e^{2W_T} \delta A_z \left(2B^2 e^{-4W_T} f^{-1} f^\prime -48 f^{-1} f^\prime- 4B^2 e^{-4W_T} W_L^\prime + 4f^{-1} f^{\prime2}W_L^\prime \right. \nonumber\\
&\left. + 8f^\prime W_L^{\prime2} -20 B^2 e^{-4W_T} +96 W_T^\prime + 8 f^{-1} f^{\prime2} W_T^\prime +16 f^\prime W_T^\prime W_L^\prime W_T^\prime  -16 f W_T^\prime W_L^{\prime2} \right. \nonumber \\
&\left.- 48 f W_L^\prime W_T^{\prime2} -32 f W_T^{\prime3} \right)=0,\label{eom_gty}\\
& \partial_r\left(e^{W_L+2W_T}\partial_r \delta V_t\right) + 8\alpha B \partial_r \delta A_z - \frac{q^2 e^{W_L}}{f}\delta V_t -\frac{iq B e^{W_L}}{f}\delta g_{ty}=0, \label{eom_vt}\\
& \partial_r\left(e^{2W_T-W_L}f \partial_r \delta A_z\right) - q^2 e^{-W_L}\delta A_z + iq \lambda e^{2W_T} \left(-4 B^2 e^{-4W_T} + 4 f^\prime W_L^\prime + 8 f^\prime W_T^\prime \right. \nonumber\\
&\left.- 8f W_T^\prime W_L^\prime -16 f W_T^{\prime2}\right)\partial_r \delta g_{ty}+ 8\alpha B \partial_r \delta V_t=0. \label{eom_az}
\end{align}
The boundary conditions on the horizon need to be derived by matching with the horizon solutions in the limit $\o\to0$. We elaborate on the derivation in appendix \ref{horizon_matching}.
The resultant boundary conditions on the horizon are given by
\begin{align}\label{bc_rh}
  \d g_{ty}(r=r_h)=0, \qquad
  \d V_t(r=r_h)=0, \qquad
  \d A_z(r=r_h)=\text{constant}.
\end{align}
The free parameters for the three fields can be chosen as horizon derivatives of $\d g_{ty}$, $\d V_t$ and horizon value of $\d A_z$. The three parameters on the horizon can be mapped to boundary values of the three fields.
We can further simplify the equations \eqref{eom_gty}-\eqref{eom_az} by considering the limit $q\to0$. Note that $\d g_{ty}$ has an opposite parity to those of $\d V_t$ and $\d A_z$, and $\d g_{ty}$ is the only field sourced on the $AdS$ boundary. Therefore, the $AdS$ boundary conditions are
\begin{align}\label{bc_rinf}
  &\delta g_{ty} \xrightarrow[]{r\to \infty} h_{ty}, \qquad {\rm others} \xrightarrow[]{r\to \infty} 0.
\end{align}
We expect the following scaling behaviors $\d g_{ty}\sim \mathcal{O}(q^0)$, and $\d V_t,\;\d A_z\sim \mathcal{O}(q)$. Defining $\d V_t=iq\d \tilde{V}_t$ and $\d A_z=iq\d \tilde{A}_z$, we can further simplify \eqref{eom_gty} through \eqref{eom_az} by keeping the leading terms in the $q$-expansion
\begin{align}\label{eom_q0}
    &\partial_r\left(e^{W_L+4W_T} \partial_r \delta g_{ty}\right)- \frac{B^2 e^{W_L}} {f(r)} \delta g_{ty}=0, \nonumber\\
& \partial_r\left(e^{W_L+2W_T}\partial_r \delta \tilde{V}_t\right) + 8\alpha B \partial_r \delta \tilde{A}_z -\frac{ B e^{W_L}}{f}\delta g_{ty}=0, \nonumber\\
& \partial_r\left(e^{2W_T-W_L}f \partial_r \delta \tilde{A}_z\right) +  \lambda e^{2W_T} \left(-4 B^2 e^{-4W_T} + 4 f^\prime W_L^\prime + 8 f^\prime W_T^\prime \right. \nonumber\\
&\left.- 8f W_T^\prime W_L^\prime -16 f W_T^{\prime2}\right)\partial_r \delta g_{ty}+ 8\alpha B \partial_r \delta \tilde{V}_t=0.
\end{align}

\section{The correlators $\langle J^t T^{ty} \rangle$ and $\langle J_5^z T^{ty}\rangle$}\label{caseI_study}

In this section, we calculate the generation of $J^t$ and $J_5^z$ as linear response to the external source $h_{ty}$. The bulk EOMs \eqref{eom_q0} will be solved under the boundary conditions \eqref{bc_rh} and \eqref{bc_rinf}. This section will be further split into two parts: an analytical study when the magnetic field is weak versus a numerical study when the value of the magnetic field is generic. In these two complementary studies, we will utilize the results of the background metric functions summarized in appendix \ref{details_background}.

In the limit $\o\to0$ and $q\to0$, the vector charge density and the axial current are (in terms of the bulk fields)
\begin{align} \label{Jt_J5z_bulk}
&J^t=\lim_{r\to \infty} \left\{e^{W_L+2W_T} \partial_r \delta V_t - \frac{e^{W_L}} {\sqrt{f(r)}} \left(B \partial_x \delta g_{ty}
\right)\log r \right\},  \nonumber \\
&J_5^z=\lim_{r\to \infty} \left\{-f(r)e^{2W_T-W_L} \partial_r \delta A_z  \right\}.
\end{align}
Near the $AdS$ boundary, the bulk fluctuations behave as
\begin{align}\label{fg_series}
&\delta g_{ty} \xrightarrow[]{r\to \infty} h_{ty}+ \frac{B^2 h_{ty}}{4r^4} \log\frac{r_h}{r} + \frac{t_{ty}}{r^4}+ \mathcal{O}(r^{-5}), \nonumber \\
&\delta \tilde V_t \xrightarrow[]{r\to \infty}  \frac{B h_{ty}}{2r^2}  \log\frac{r_h}{r} + \frac{v_t^2}{r^2} + \mathcal{O}(r^{-3}\log r), \nonumber \\
&\delta \tilde A_z \xrightarrow[]{r\to \infty} \frac{a_z^2}{r^2}+ \mathcal{O}(r^{-3}).
\end{align}
So, the vector charge density and axial current for the boundary theory are
\begin{align}\label{general}
J^t= -2 v_t^2 iq - B iq h_{ty} \left(\log r_h+ \frac{1}{2}\right), \qquad \qquad J_5^z= 2 iq a_z^2.
\end{align}
Below we solve for $J^t$ and $J_5^z$ perturbatively in $B$ and also numerically for generic $B$.

\subsection{Weak magnetic field: a perturbative study} \label{weak_B_study}

When the magnetic field $B$ is weak, the bulk fluctuations are expandable
\begin{align}
&\delta g_{ty}= \delta g_{ty}^{[0]}+ \epsilon^2\delta g_{ty}^{[2]}+ \cdots,\nonumber \\
&\delta \tilde V_t= \epsilon\delta \tilde V_t^{[1]}+ \epsilon^3 \delta \tilde V_t^{[3]}+ \cdots, \nonumber \\
&\delta \tilde A_z= \delta \tilde A_z^{[0]} + \epsilon^2 \delta \tilde A_z^{[2]}+ \cdots,
\end{align}
where $\epsilon\sim B$. At the lowest order $\mathcal{O}(\epsilon^0)$, first we have
\begin{align}
0=\partial_r \left(r^5 \partial_r \delta g_{ty}^{[0]}\right) \Longrightarrow \delta g_{ty}^{[0]}= \left(1-\frac{r_h^4}{r^4}\right)h_{ty}.
\end{align}
Then, we have
\begin{align}
0=\partial_r\left((r^3-r_h^4/r)\partial_r \delta \tilde A_z^{[0]}\right)+ \frac{48 \lambda r_h^4}{r^2} \partial_r \delta g_{ty}^{[0]},
\end{align}
whose solution is
\begin{align}
\delta \tilde A_z^{[0]}= 8\lambda \left( \frac{r_h^4}{r^4}+ 2\log \frac{r^2+r_h^2}{r^2} \right) h_{ty}.
\end{align}

At the first order $\mathcal{O}(\epsilon^1)$,
\begin{align}
0= \partial_r\left(r^3 \partial_r \delta \tilde V_t^{[1]}\right) +8\alpha B \partial_r \delta \tilde A_z^{[0]}- \frac{B}{r}h_{ty},
\end{align}
which is solved by
\begin{align}
\delta \tilde V_t^{[1]} &= B h_{ty} \left\{ \frac{\log(r_h/r)}{2r^2} - \frac{128(\log 2)\alpha\lambda}{r^2}+ \frac{32\alpha \lambda}{3r^6}(r_h^4-r^4) + 64 \alpha\lambda \left(\frac{1}{r^2}+ \frac{1}{r_h^2} \right)\log\frac{r^2+r_h^2}{r^2} \right\}\nonumber\\
&\xrightarrow[]{r\to \infty} \left\{- \frac{\log r}{2r^2} +\frac{1}{r^2} \left(\frac{1}{2}\log r_h + \frac{160}{3} \alpha \lambda -128 \log 2 \; \alpha\lambda\right) \right\} B h_{ty} + \mathcal{O}(r^{-3}),
\end{align}

At the second order $\mathcal{O}(\epsilon^2)$:
\begin{align}
0= \partial_r\left(r^5 \partial_r \delta g_{ty}^{[2]}\right)- \frac{B^2}{r}h_{ty}+ 8r_h^4 W_T^{(2)\prime}(r) h_{ty}.
\end{align}
Here, we would like to remind that $W_T^{(2)}$ is obtained in \eqref{WT2}.
The solution for $\delta g_{ty}^{[2]}$ would be
\begin{align}
\delta g_{ty}^{[2]}(r)=- B^2 h_{ty}\int_r^\infty \frac{\log x}{x^5}dx + 8r_h^4 h_{ty} \int_r^\infty \frac{W_T^{(2)}(x)}{x^5}dx + \frac{C}{r^4},
\end{align}
where the integration constant $C$ is fixed as
\begin{align}
\delta g_{ty}^{[2]}(r=r_h)=0 \Longrightarrow C&= B^2r_h^4 h_{ty}\int_{r_h}^\infty \frac{\log x} {x^5} dx - 8r_h^8 h_{ty}\int_{r_h}^\infty \frac{W_T^{(2)}(x)}{x^5}dx \nonumber\\
&= \frac{1}{24}B^2 h_{ty} (1+6\log r_h).
\end{align}
So,
\begin{align}
\partial_r \delta g_{ty}^{[2]}(r)= h_{ty} \left\{\frac{B^2\log r}{r^5} - \frac{8r_h^4}{r^5} W_T^{(2)}(r)- \frac{1}{6r^5}B^2(1+6\log r_h)\right\}.
\end{align}
The equation for $\delta \tilde A_z^{[2]}$ is
\begin{align}
0=\partial_r\left[(r^3-r_h^4/r)\partial_r \delta \tilde A_z^{[2]}\right]+ \alpha H_1(r)+ \alpha^2 \lambda H_2(r)+ \lambda H_3(r),
\end{align}
where
\begin{align}
&H_1(r)= h_{ty} \left(-\frac{4B^2}{r^3}- \frac{8B^2\log(r_h/r)}{r^3}\right), \nonumber \\
&H_2(r)=h_{ty}\left[-\frac{512 B^2 r_h^4}{r^7}+ \frac{512B^2(12\log2-5)}{3r^3} -\frac{1024 B^2 }{r^3}\log \left(1+\frac{r_h^2}{r^2}\right)\right], \nonumber \\
&H_3(r)= h_{ty} \left\{- \frac{80 B^2 r_h^4}{3r^7} + \frac{16B^2 r_h^2}{3r^3(r^2+r_h^2)} + \frac{64r_h^2(r_h^4-r_p^4)}{r(r^2+r_h^2)}\left(\frac{1}{r^2+r_h^2}+ \frac{1}{r^2} \right) \right. \nonumber\\
&\qquad \qquad \qquad \left.+ \left[\frac{32B^2 r_h^2}{3r(r^2+r_h^2)^2} + \frac{32B^2 r_h^2}{3r^3 (r^2+r_h^2)} - \frac{48 B^2 r_h^4}{r^7}\right] \log\frac{r_h}{r} - \frac{768r_h^8 W_T^{(2)}}{r^7} \right. \nonumber \\
&\qquad \qquad \qquad \left. +32 r_h^2 W_T^{(2)\prime} \left[\frac{7r_h^2(r_h^6-r^6)} {r^6} + \frac{4(r_h^4-r^4)}{r^2(r^2+r_h^2)}\right]\right\}.
\end{align}
The solution as well as near-boundary expansion for $\delta \tilde A_z^{[2]}$  are
\begin{align}
\delta \tilde A_z^{[2]}(r)&=\int_r^\infty \frac{dx}{x^3-r_h^4/x} \int_{r_h}^x \left[\alpha H_1(y) + \alpha^2 \lambda H_2(y) + \lambda H_3(y)\right]dy  \nonumber \\
&\xrightarrow[]{r\to \infty} -\frac{4\lambda h_{ty}(B^2-6r_h^4+6r_p^4)}{3r_h^2r^2}+ \mathcal{O}(r^{-3}).
\end{align}

Up to $\mathcal{O}(B^2)$, the vector charge density and axial current on the boundary are
\begin{align}
&J^t= \left[\frac{64}{3}(12\log2-5)\alpha\lambda - \log r_h - \frac{1}{2}\right]B iq h_{ty}+ \cdots, \nonumber \\
&J_5^z=32\lambda r_h^2 iq h_{ty} - \frac{8\lambda}{3r_h^2}\left(B^2-6r_h^4+6r_p^4\right) iq h_{ty} + \cdots. \label{Jt_J5z_hty}
\end{align}
The transport coefficients $\xi$ and $\sigma$ are
\begin{align}\label{xs}
&\xi= \frac{128}{3}(12\log2-5)\alpha\lambda - 2\log r_h -1+ \mathcal{O}(B/T^2), \nonumber \\
&\sigma= r_h^2 \left[64\l - \frac{32\lambda B}{3r_h^2}+ \mathcal{O}(B^2/T^4) \right],
\end{align}
where we have substituted the perturbative expression \eqref{rp_perturb} for $r_p$.

The transport coefficient $\x$ contains both non-anomalous contribution and anomalous contribution proportional to $\a\l$. The non-anomalous contribution is consistent with the prediction of the non-anomalous magnetohydrodynamics (MHD) \cite{Hernandez:2017mch}, which for a neutral plasma has\footnote{In comparison with \cite{Hernandez:2017mch}, we have included an overall factor $2$ given that the fluid's vorticity defined in \eqref{vorticity} is half of that of \cite{Hernandez:2017mch}.}
\begin{align}\label{mhd_N}
  \Delta J^t=2 \left(M_{\O,\m}{\vec B}\cdot{\vec \O}-2p_{,B^2}{\vec B}\cdot{\vec \O} \right).
\end{align}
with $M_\O \equiv\frac{\pd \mathcal{F}}{\pd (B\cdot\O)}$ being the magneto-vortical susceptibility and $2p_{,B^2}\equiv2\frac{\pd p}{\pd (B^2)}$ being the magnetic susceptibility. Here, $p$ is the pressure and $\mathcal{F}$ is the free energy density \cite{Hernandez:2017mch}. Note that in \eqref{mhd_N} we ignored terms nonlinear in $\O$. Both susceptibilities $M_\O$ and $2p_{,B^2}$ can be calculated independently. In the weak $B$ field limit, the magnetic susceptibility $2p_{,B^2}$ can be calculated from the perturbative background we already obtain in appendix \ref{details_background}. While the magneto-vortical susceptibility $M_\O$ vanishes for a neutral plasma by charge conjugation symmetry, $M_{\O,\m}\equiv \partial M_\O/\partial \mu$ does not. The quantity $M_{\O,\m}$ can even be calculated in a charged plasma at $B=0$. In appendix \ref{app_suscept}, we calculate both susceptibilities with the following results
\begin{align}\label{suscept}
  2p_{,B^2}=\log r_h,\quad\quad M_{\O,\m}=-\frac{1}{2}.
\end{align}
The negative value of $M_{\O,\m}$ is consistent with the fact that spin-vorticity coupling lowers/raises energy of particle/anti-particle.
Clearly, \eqref{mhd_N} and \eqref{suscept} are in perfect agreement with \eqref{Jt_J5z_hty}.

The dependence $\log r_h$ may look odd at the first sight. To restore unit, we should use the replacement $\log r_h\to\log(r_h L)$. In fact, this transport coefficient is scheme dependent. The appearance of the $AdS$ radius $L$ comes from the fact we use $1/L$ as our renormalisation scale \cite{Fuini:2015hba}. Other physically significant renormalisation scale could be used, which could alter this term \cite{Fuini:2015hba}. It is also interesting to note that the scheme dependence is related conformal anomaly. In fact, a different scheme would correspond to adding a finite counter term as
\begin{equation}\label{counter_finite}
\D S_{\rm c.t.}=-\frac{1}{2\kappa^2} \int d^4x \sqrt{-\gamma}\left(\frac{a}{4}\left(F^V\right)_{\mu\nu} \left(F^V\right)^{\mu\nu}\right).
\end{equation}
Such a counter term would give the following contribution to the vector current
\begin{align}\label{j_finite}
  \D J^\m=-\frac{a}{2\kappa^2}\sqrt{-\g}\nabla_\n F^{\m\n}.
\end{align}
In the presence of $h_{ty}(x)$, we can easily obtain $\D J^t\sim a {\vec B}\cdot{\vec \O}$. Therefore the combination $M_{\O,\m}-2p_{,B^2}$ can be shifted by a constant. Note that the scheme dependence of the vector charge density is absent in a free theory.

The anomalous contribution is proportional to $\l\a B$. The EOM \eqref{eom_q0} suggests the following chain of responses: $\d A_z\sim \mathcal{O}(\l)$ is induced in response to vorticity and then backreaction of $\d A_z$ to $\d V^t$ gives $J^t\sim \mathcal{O}(\l\a B)$. This corresponds to the backreaction of $J_5^z$ generated by CVE to $J^t$ on the field theory side.
A possible $\alpha^2B^2$-term would emerge at the next order $\mathcal{O}(B^2)$. In this case, \eqref{eom_q0} suggests the following chain of responses: a non-anomalous contribution to $J^t$ is generated by magneto-vortical coupling. Then $J_5^z$ is induced by CSE. The backreaction of $J_5^z$ to $J^t$ would give the $O(\alpha^2B^2)$ contribution. However, the above reasoning is not quite accurate. As we show below, in fact CSE is not generated in the presence of $J^t$. Nevertheless the bulk profile of $\d A_z$ does backreact to $\d V_t$ to give $\alpha^2B^2$ correction to $J^t$.

In contrast to $\x$, the transport coefficient $\s$ is scheme independent. We can add analogous counter term as \eqref{counter_finite}. It would not contribute to $J_5^z$ in the absence of background axial gauge field. The structure of $\s$ is relatively simple. Aside from the $T^2$-correction to the CVE (i.e., the first piece in $\s$), $\s$ encodes correction to chiral vortical coefficient from $B$. However, there is no contribution proportional to $\a$. In other words, no CSE is seen despite the generation of $J^t$. This is in contrast to the naive expectation from CSE
\begin{align}\label{cse}
  J_5^z=8\a B\frac{J^t}{\chi},
\end{align}
with $\chi$ being vector charge susceptibility.
In fact, from the holographic model, the absence of CSE holds more generally: if we integrate \eqref{eom_az} from the horizon to an arbitrary $r$, we obtain (with $\l=0$)
\begin{align}\label{eom_az_int}
e^{2W_T-W_L}f\partial_r \delta \tilde A_z+ 8 \alpha B \delta \tilde V_t(r)=0,
\end{align}
where the horizon boundary conditions \eqref{bc_rh} have been utilised to fix the integration constant. Taking $r\to\infty$ and noting $\tilde{V}_t(r\to\infty)=0$, we have
\begin{align} \label{J5z_lambda=0}
J_5^z=0,\qquad {\rm when} \qquad \lambda=0.
\end{align}
Following \cite{Gynther:2010ed}, we should identify the difference of $\d V_t$ on the boundary and on the horizon as the vector chemical potential
\begin{align}\label{mu_def}
  \m=\d V_t(r=\infty)-\d V_t(r=r_h).
\end{align}
It then follows naturally from \eqref{eom_az_int} that $J_5^z=8\a B\m$. In our case, we have $\m=0$ but $J^t\ne0$.

To understand the physical difference between $J^t$ and $\m$, we note $\m$ ($-\m$) is the extra energy cost to create one unit of particle (anti-particle), but $J^t$ depends on actual distribution of particles and anti-particles. In our case $\m=0$ implies that it costs the same energy to create both particle and anti-particle. Indeed, we can view $\d V_t$ as a zero mode as it vanishes both on the horizon and on the boundary, which supports the picture of vanishing energy cost for creating particle. Since the state is obtained by the adiabatic limit, it means $J^t$ is generated dynamically\footnote{Interestingly, similar situations have been found for equilibrium state \cite{Tatsumi:2014wka,Bu:2018trt}.}. The CSE seems to be only sensitive to the energy difference $\m$, not the charge density. The absence of CSE can also been understood from the scheme independence $J^z_5$: unlike $J^t$, $J^z_5$ is unaffected by the choice of scheme from \eqref{Jt_J5z_bulk}. This makes natural for $J_5^z$ to depend on $\m$, rather than on $J^t$.

\subsection{Generic magnetic field: a numerical study}

For generic value of $B$, we will solve the fluctuation EOMs \eqref{eom_q0} through the shooting technique. First, we find out the near-horizon solution:
\begin{align}
&\delta g_{ty}= \delta g_{ty}^1 (r-r_h)+ \delta g_{ty}^2 (r-r_h)^2+ \cdots, \nonumber \\
&\delta \tilde V_t= \delta \tilde V_t^1 (r-r_h) + \delta \tilde V_t^2 (r-r_h)^2+ \cdots, \nonumber \\
&\delta \tilde A_z= \delta \tilde A_z^0+ \delta \tilde A_z^1 (r-r_h) + \delta \tilde A_z^2 (r-r_h)^2 + \cdots,
\end{align}
where, thanks to the horizon condition \eqref{bc_rh}, only $ \delta g_{ty}^1$, $\delta \tilde V_t^1$ and $\delta \tilde A_z^0$ are undetermined. Then, we will choose a reasonable value of $\delta g_{ty}^1$ (corresponds to turning on a specific source $h_{ty}$) and finely tune $\delta \tilde V_t^1, \delta \tilde A_z^0$ until $\delta \tilde V_t(r=\infty)= \delta \tilde A_z(r=\infty)=0$ are satisfied. From the numerical solution, we can read off the expectation values of $J^t$ and $J_5^z$, as response to the source $h_{ty}$ only. In practical numerics, we set the horizon data $\delta g_{ty}^1=-1$.

However, there is one problem in the procedure mentioned above.
%
%
%
Since we intend to solve the background EOMs \eqref{f}-\eqref{Wl} using the initial conditions \eqref{UVW_horizon} and \eqref{UVW_horizon1}, we should be careful in solving the fluctuation EOMs \eqref{eom_q0}. More precisely, the correct solutions are
\begin{align}
\delta g_{ty}= \frac{\delta g_{ty}^*}{\sqrt{v(b)}}, \qquad \delta V_t= \delta V_t^*, \qquad \delta A_z= \frac{\delta A_z^*}{\sqrt{w(b)}},
\end{align}
where the stared functions $\delta g_{ty}^*$, $\delta V_t^*$ and $\delta A_z^*$ are solved from \eqref{eom_q0} using the ``incorrect'' numerical background metric functions, as discussed in appendix \ref{details_background}. Adapted to the tilde variables, we have
\begin{align}
\frac{\delta \tilde V_t}{B} = \frac{\delta V_t}{iq B}=v^{3/2}(b) \delta \tilde V_t^*, \qquad \qquad \delta \tilde A_z \equiv \frac{\delta A_z}{iq}= \sqrt{\frac{v(b)}{w(b)}} \delta \tilde A_z^*.
\end{align}
Here, for the sake of numerical calculation, we have further re-scaled the $\delta \tilde V_t$ of \eqref{eom_q0} by a factor of $B$.

For convenience, we set $r_h=1$ in our numerical calculations. So, the dimensionful quantities ($J^t, J_5^z$ etc) to be plotted in sections \ref{medium_effect} and \ref{anomalou_effect} should be understood as in units of proper powers of $r_h$.

%

\subsubsection{Non-anomalous effects: $\alpha=\lambda=0$} \label{medium_effect}

\begin{figure}[htbp]
\centering
\includegraphics[width=0.48\textwidth]{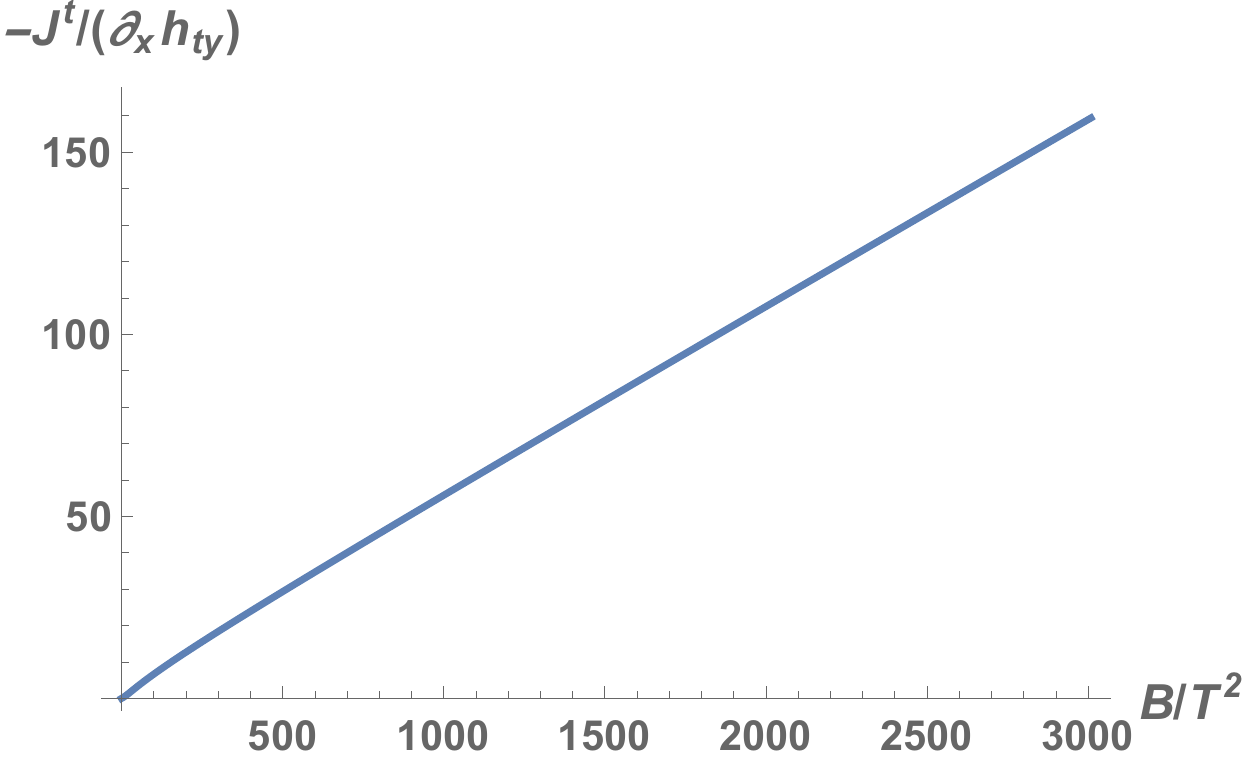}
\includegraphics[width=0.48\textwidth]{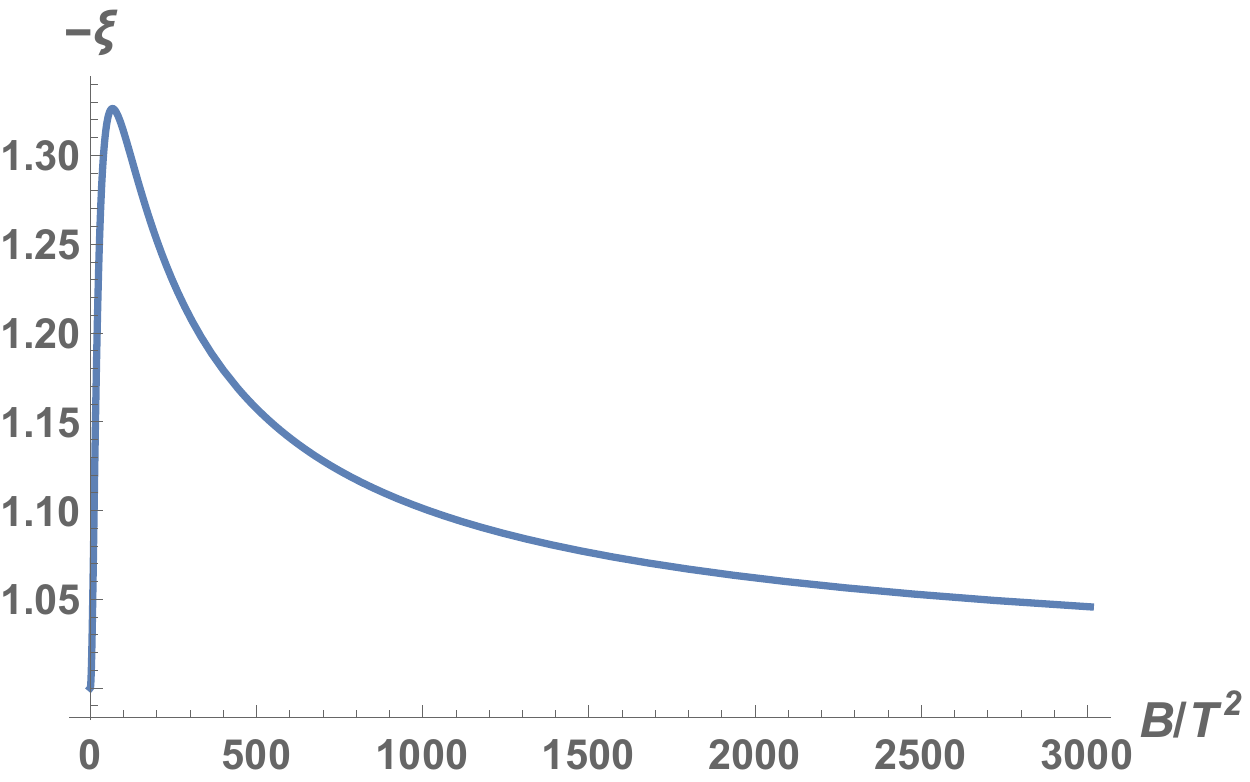}
\caption{The correlator $\langle J^t T^{ty}\rangle/(iq)=J^t/\partial_x h_{ty}$ in unit of $r_h^2$ (left) and the transport coefficient $\xi$ (right) as a function of $B/T^2$ when $\alpha=\lambda=0$.}
\label{Jtoverhty_xi_alpha0_lambda0}
\end{figure}

When the chiral anomaly and gravitational anomaly are turned off (i.e., $\alpha=\lambda=0$), the transport properties of the magnetized plasma get non-anomalous contributions from the medium only. The medium effects are not covered by the study of \cite{Hattori:2016njk} since the calculations therein are essentially based on the vacuum state. In this situation, as we discussed in the previous section we only see a dynamically generated vector charge density $J^t$, whereas the correlator $\langle J^z_5 T^{ty}\rangle/(iq)$ (and thus $J_5^z$) vanishes identically as seen from \eqref{J5z_lambda=0}. In Figure \ref{Jtoverhty_xi_alpha0_lambda0}, we show the correlator $\langle J^t T^{ty}\rangle/(iq)$ and the transport coefficient $\xi$ as a function of $B/T^2$. For the purpose of probing the strong magnetic field limit, we have improved our numerical calculations and generate plots up to $B/T^2\sim 3000$.

\begin{figure}[htbp]
\centering
\includegraphics[width=0.48\textwidth]{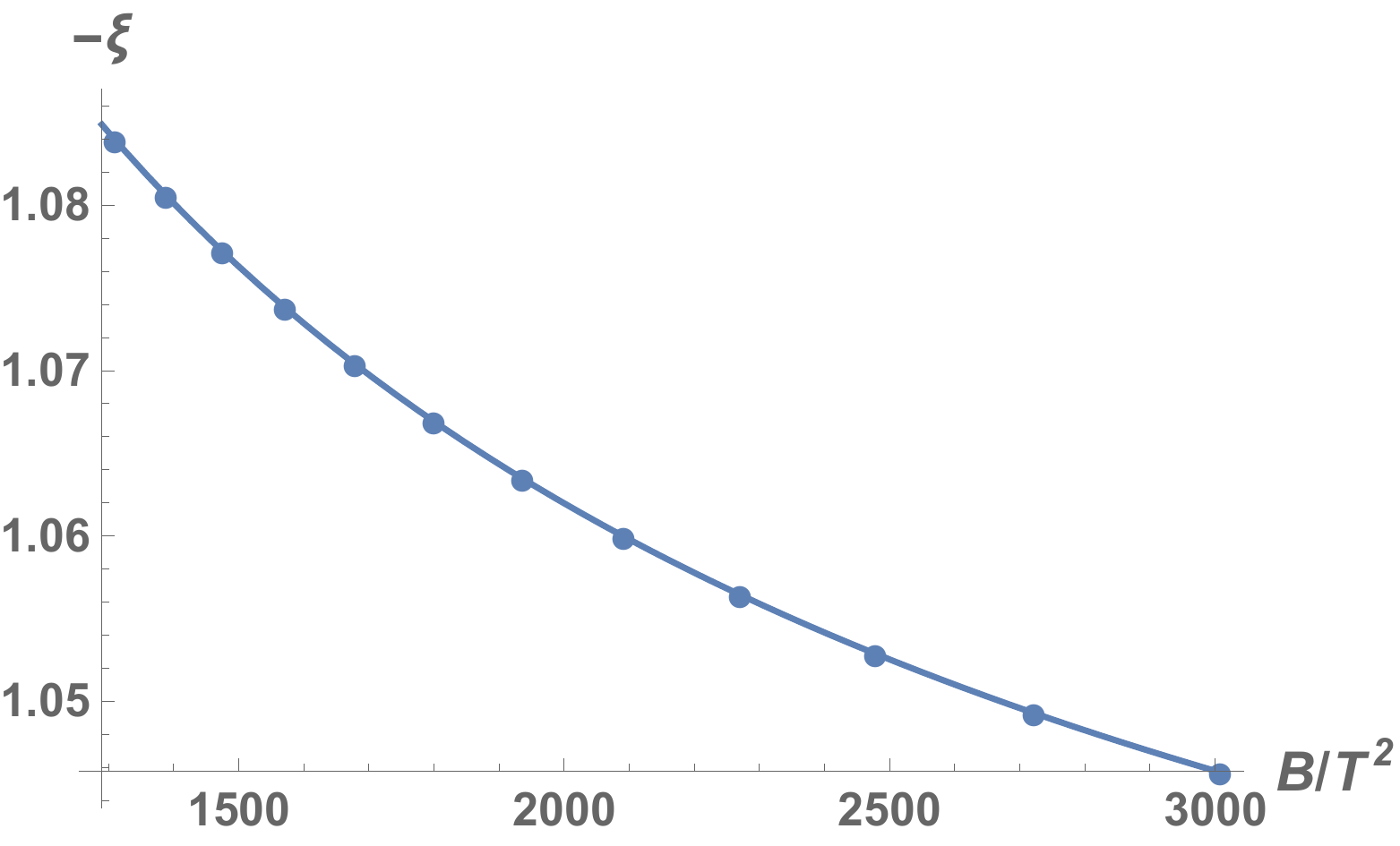}
\caption{In the strong $B$ limit, the numerical result of $\xi$ (dots) is fitted by the function \eqref{fit_xi} when $\alpha=\lambda=0$.}
\label{xi_alpha0_lambda0_fitting}
\end{figure}

From the left panel of Figure \ref{Jtoverhty_xi_alpha0_lambda0}, it {\it seemingly} implies a quasi-linear growth for $\langle J^t T^{ty}\rangle/(iq)$ as $B/T^2$ is increased, which as we will show is inaccurate.
%
%
The right panel of Figure \ref{Jtoverhty_xi_alpha0_lambda0} reveals more information: $-\x$ approaches $1$ from above. In Figure \ref{xi_alpha0_lambda0_fitting}, we fit our numerical result for $\xi$ in the strong magnetic field limit by the following function:
\begin{align} \label{fit_xi}
-\x \simeq 1.0001779 + 32.692107 \frac{\log(B/T^2)}{B/T^2} + \frac{124.89109}{B/T^2}.
\end{align}
It is tempting to conclude that $-\xi\to1$ asymptotically. The correction in \eqref{fit_xi} can be understood as the $v_t^2$ term in the general expression \eqref{general} by noting that $r_h=1$ in our numerical results. 



\subsubsection{Anomalous effects: $\lambda \neq 0$} \label{anomalou_effect}

We now turn to anomalous contributions to the transport properties of the magnetized plasma. While both anomaly coefficients $\a, \l$ are fixed for a specific QFT on the boundary, we here take a phenomenological viewpoint and think of $\a,\l$ as free parameters. First of all, taking $\l =0$ will kill $J_5^z$ completely, as seen from the bulk EOMs \eqref{eom_q0}. Thus, our representative choices for the anomaly coefficients are:
\begin{align} \label{alpha_lambda}
(\a,\l)=(0,1/50), \; (1/20, 1/50), \; (1/20,1/20).
\end{align}

\begin{figure}[htbp]
\centering
\includegraphics[width=0.48\textwidth]{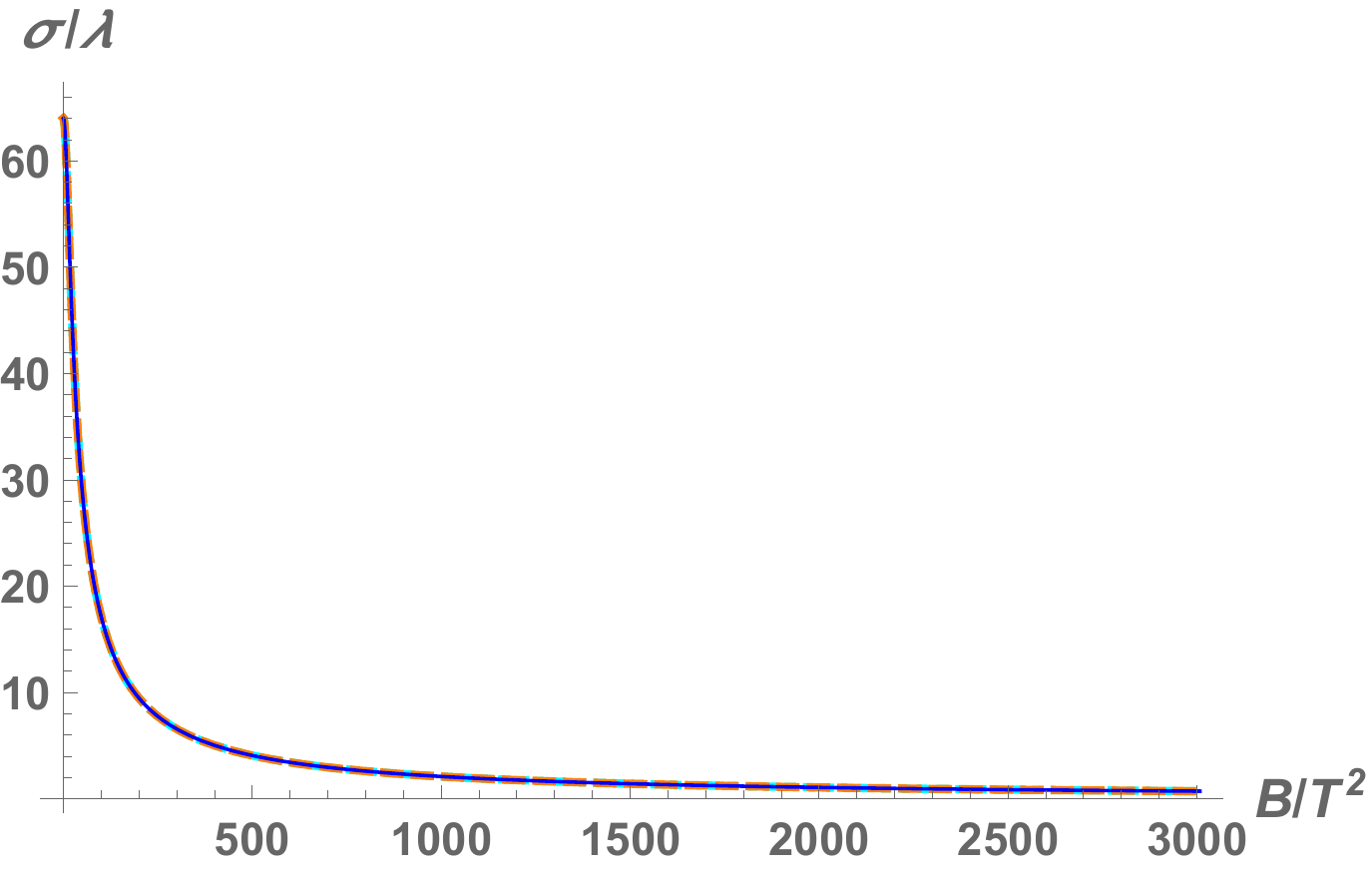}
\includegraphics[width=0.48\textwidth]{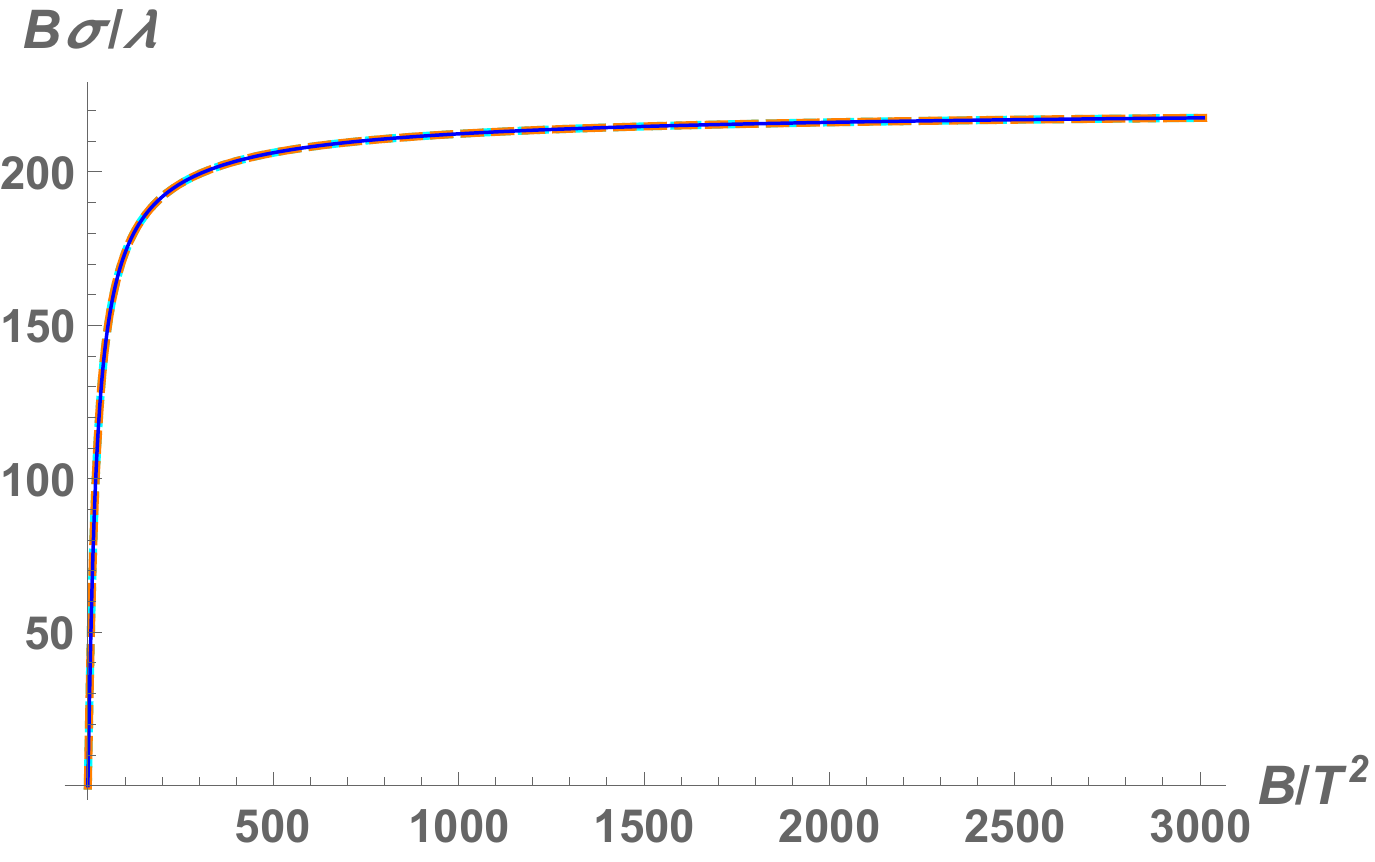}
\caption{Left: The CVE conductivity $\s$ in unit of $r_h^2$ (with anomaly coefficient $\l$ factorized out) as a function of $B/T^2$. Right: $B\s/\l$ in unit of $r_h^4$, demonstrating the asymptotic behavior for $\s$ at strong magnetic field. Here, in each plot the three curves (dashed orange, solid blue, dotted green) corresponding to the different choice in \eqref{alpha_lambda} perfectly collapse into a single one.}
\label{sigma_alpha_lambda}
\end{figure}

We begin with the fate of the CVE conductivity $\s$. First, the last equation of \eqref{eom_q0} could be formally integrated from the horizon to the $AdS$ boundary, yielding:
\begin{align}
J_5^z= \sigma \O + 8\a \m B,
\end{align}
where $\m$ is defined in \eqref{mu_def}. Here, we stress that the CVE conductivity $\s$ depends on $\l$ linearly and is independent of $\a$. In the left panel of Figure \ref{sigma_alpha_lambda}, we plot the CVE conductivity $\sigma$ as a function of $B/T^2$, taking all choices for $\a,\l$ from \eqref{alpha_lambda}. From the plot, we obviously see perfect overlapping of different curves, confirming our claim that $\s$ linearly depends on $\l$ only. Intriguingly, the magneto-vortical coupling effect tends to suppress the CVE conductivity and eventually renders it to vanish at large magnetic field. Asymptotically, $\sigma \sim B^{-1}$ as demonstrated in the right panel of Figure \ref{sigma_alpha_lambda}. Similar suppression effects due to quark mass \cite{Lin:2018aon,Ji:2019pxx} and spacetime curvature \cite{Flachi:2017vlp} are also seen. Our findings are in contrast to the proposal of \cite{Hattori:2016njk} that the magneto-vortical coupling (through chiral anomaly) generates a linear in $B$ term to $\s$ when $B/T^2$ becomes very large.

\begin{figure}[htbp]
\centering
\includegraphics[width=0.48\textwidth]{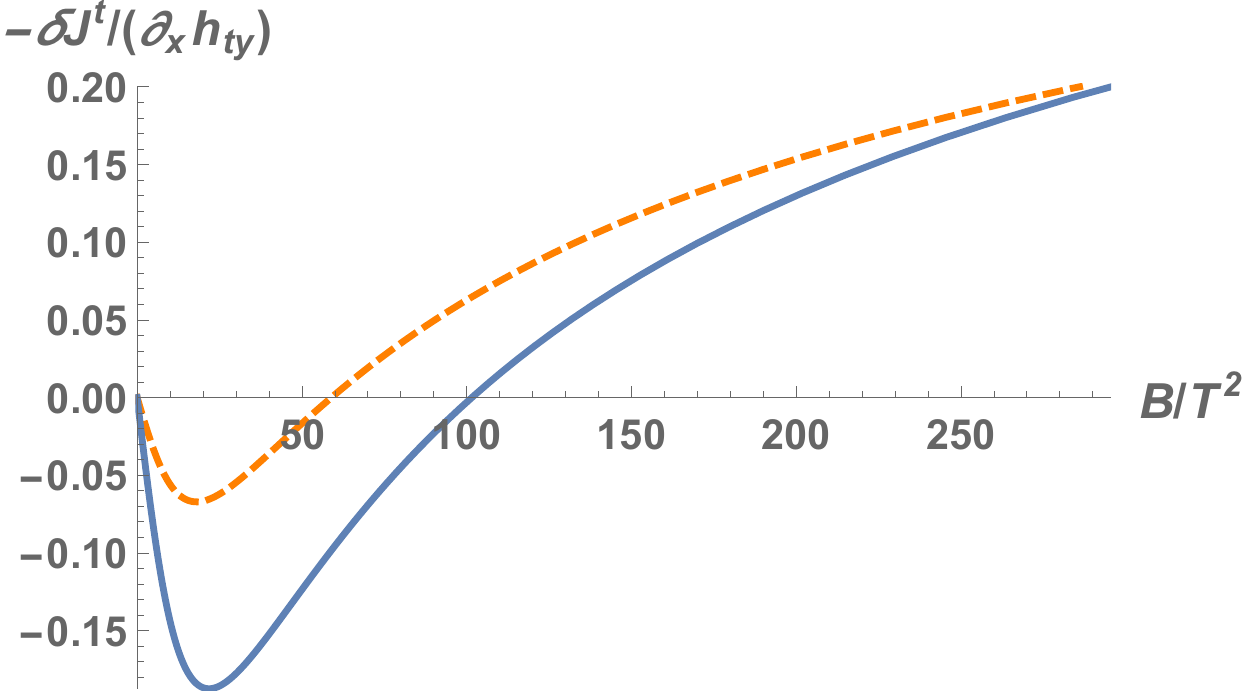}
\includegraphics[width=0.48\textwidth]{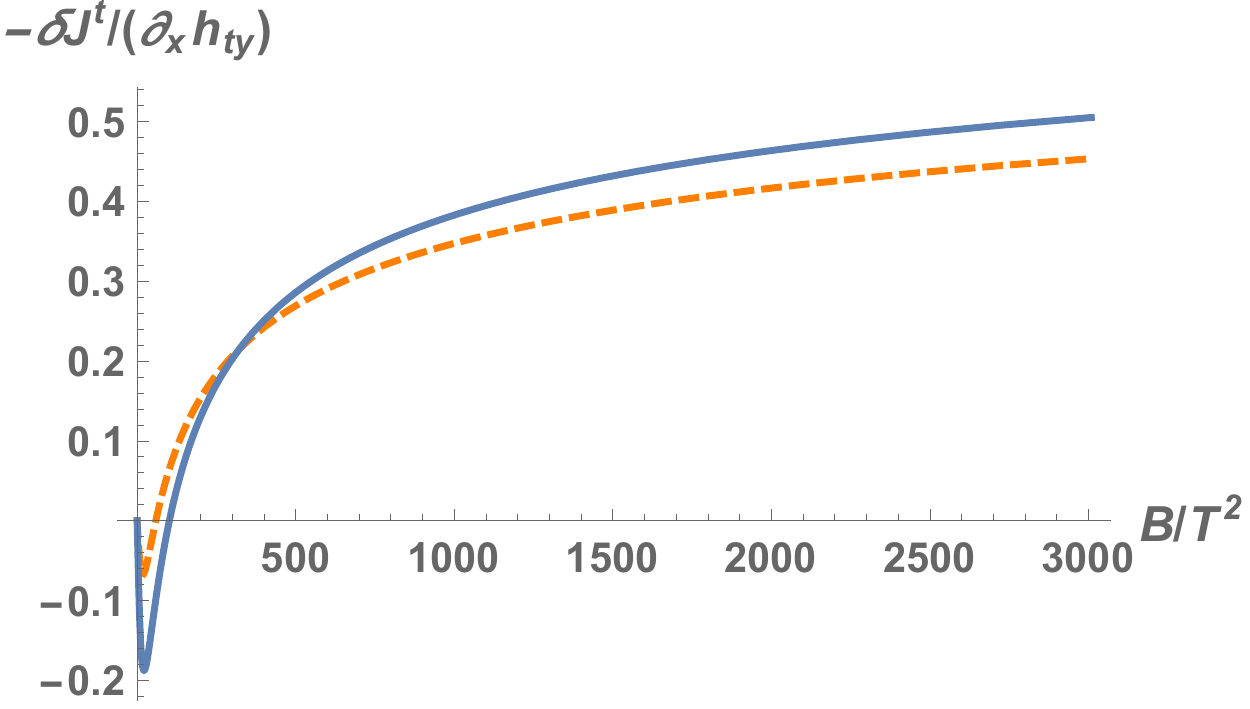}
\caption{The anomalous correction $\langle \delta J^t T^{ty}\rangle/(iq)$ in unit of $r_h^2$ as a function of $B/T^2$:  $\alpha=1/20, \lambda=1/50$ for dashed curve and $\alpha=1/20, \lambda=1/20$ for solid one.}
\label{delJtoverhty_alpha_lambda}
\end{figure}

\begin{figure}[htbp]
\centering
\includegraphics[width=0.48\textwidth]{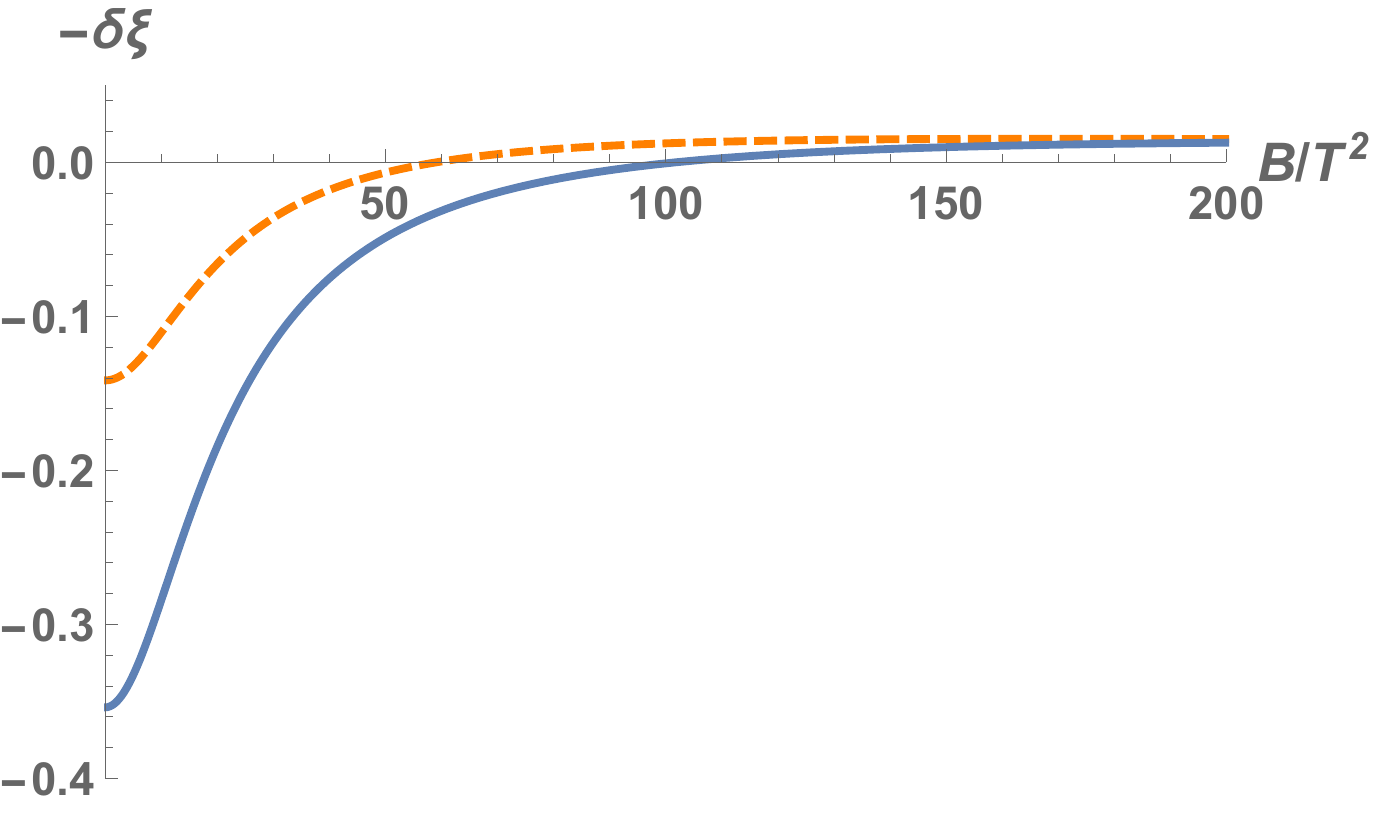}
\includegraphics[width=0.48\textwidth]{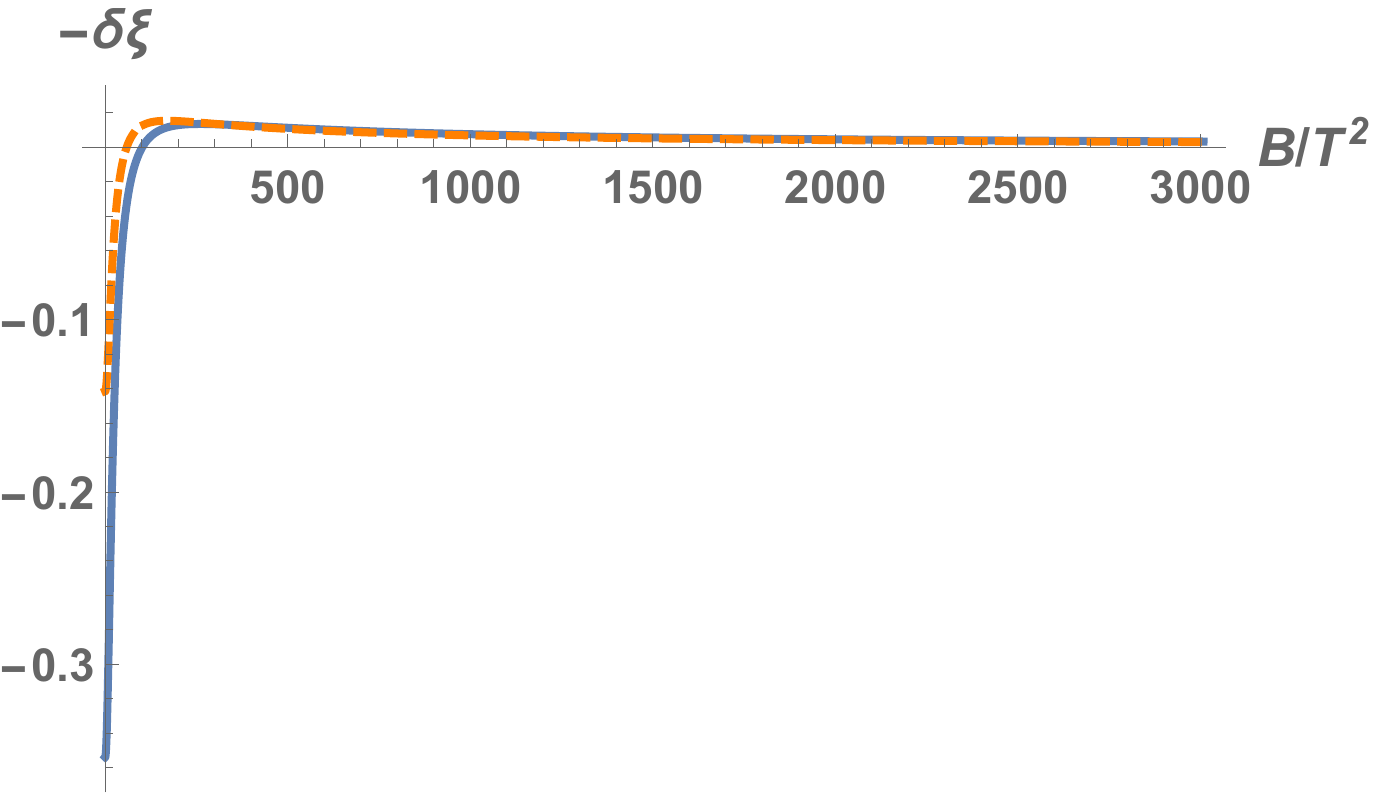}
\caption{The anomalous correction $\delta \xi$ as a function of $B/T^2$: $\alpha=1/20, \lambda=1/50$ for the dashed curve and $\alpha=1/20, \lambda=1/20$ for the solid one. }
\label{delxi_alpha_lambda}
\end{figure}

Next, we consider anomalous contributions to the generation of the vector charge density $J^t$ and the transport coefficient $\xi$. Given that $J^t$ always gets non-anomalous contribution, we find it more transparent to consider
\begin{align}
\delta J^t \equiv J^t- J^t|_{\a=\l=0}, \qquad \qquad  \delta \xi \equiv \xi- \xi|_{\a=\l=0}.
\end{align}
In accord with the different choices for $\a,\l$ as made in \eqref{alpha_lambda}, we show anomalous corrections $\langle \delta J^t T^{ty}\rangle/(iq)$ and $\delta \xi$ in Figures \ref{delJtoverhty_alpha_lambda} and \ref{delxi_alpha_lambda}, respectively. From the second equation of \eqref{eom_q0}, it is clear that taking $\a=0$ makes anomalous contributions $\delta J^t$ and $\delta \xi$ to be zero. So, for each panel of Figures \ref{delJtoverhty_alpha_lambda} and \ref{delxi_alpha_lambda}, we have only two non-trivial curves.

\begin{figure}[htbp]
\centering
\includegraphics[width=0.48\textwidth]{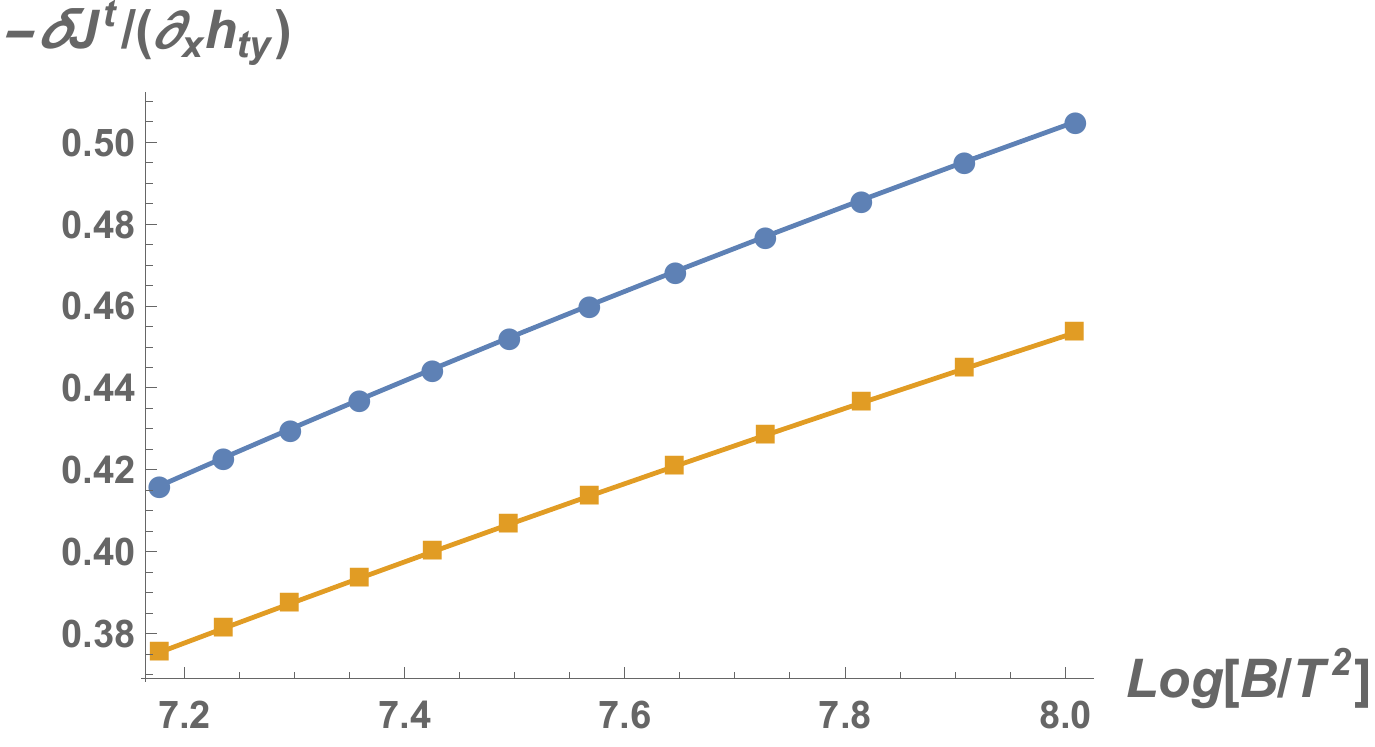}
\caption{The anomalous correction $\langle \delta J^t T^{ty}\rangle/(iq)$ in unit of $r_h^2$ as a function of $\log (B/T^2)$: $\alpha=1/20, \lambda=1/50$ for the squared points and $\alpha=1/20, \lambda=1/20$ for the circled points.}
\label{delJtoverhty_LogB_alpha_lambda}
\end{figure}

From Figure \ref{delJtoverhty_alpha_lambda}, we read that the anomalous contribution to vector charge $\d J^t$ has opposite sign from the their non-anomalous counterpart for weak magnetic field.
As the magnetic field becomes stronger, $\d J^t$ changes sign and continue to grow mildly at large $B/T^2$. 
%
More precisely, the numerical results at large $B$ imply the following asymptotic behaviors for the anomalous corrections:
\begin{align}\label{logB}
\delta J^t \simeq \log B, \qquad \delta \xi \simeq \frac{\log B}{B},
\end{align}
which are clearly confirmed by the plots of Figure \ref{delJtoverhty_LogB_alpha_lambda}.
It is worth noting that from \eqref{fit_xi} and \eqref{logB} the large $B$ limit of $\x$ is dominated by the non-anomalous medium contribution.


%
%
%
%
%
%
%
%
%

\section{Thermal Hall effect and thermal axial magnetic effect} \label{hydro_analysis}

In the previous section, we have obtained the following transport coefficients
\begin{align}\label{jtj5z2}
  &J^t= \xi(B,T)(\vec B\cdot \vec \O),\nonumber\\
  &J_5^z=\s(B,T)\O.
\end{align}
By Onsager relation, \eqref{jtj5z2} gives rise to
\begin{align}\label{thermal}
  &T^{ty}= \frac{1}{2}\xi(-B,T)E_xB,\nonumber\\
  &T^{ty}=\frac{1}{2}\s(-B,T)B_{5y}.
\end{align}
In a neutral plasma, \eqref{thermal} corresponds to the generation of thermal current by transverse electric field and transverse axial magnetic field, which we coin thermal Hall effect and thermal axial magnetic effect, respectively. Below we will derive \eqref{thermal} more rigorously using the time-reversal symmetry. We start with the following correlators for responses to $h_{ty}$:
\begin{align}\label{gty_response}
  &\lag J^t(q)T^{ty}(-q)\rag=\frac{iq\xi B}{2}, \nonumber\\
  &\lag J^z_5(q)T^{ty}(-q)\rag=\frac{iq \s}{2},
\end{align}
with the correlators in \eqref{gty_response} being the limit $\o\to0$ of the following retarded correlator:
\begin{align}\label{R_def}
  \lag O_a(\o,q)O_b(-\o,-q)\rag&=\int\frac{dtd^3x}{(2\pi)^4}e^{-i\o t+i\vec{q}\cdot\vec{x}}G^R_{ab}(t,x,B)\nonumber\\
  &=\int\frac{dtd^3x}{(2\pi)^4}(-i)e^{-i\o t+i\vec{q}\cdot\vec{x}}\theta(t)\lag[O_a(t,x),O_b(0)]\rag.
\end{align}
By time-reversal symmetry, we can obtain the transposed correlators by \cite{Kovtun:2012rj}
\begin{align}\label{time-reversal}
  & G^R_{ab}(t,x,B)=\gamma_a\gamma_bG^R_{ba}(t,-x,-B) \nonumber\\
  \Rightarrow & \lag O_a(\o,q)O_b(-\o,-q)\rag_B=\gamma_a\gamma_b\lag O_b(\o,-q)O_a(-\o,q)\rag_{-B}
\end{align}
with $\gamma_{a}=\pm1$ corresponding to eigenvalues of operator $O_a$ under time-reversal. Note that $B$ flips sign under time-reversal, which we indicate in the subscript. For the operators of our interest ${T^{ty},J^t,J_5^z}$, we have $\gamma={-,+,+,}$ respectively. Therefore, \eqref{gty_response} and \eqref{time-reversal} give the following transposed correlators (with the direction of $B$ reversed and $\o=0$)
%
\begin{align}\label{cross_response}
  &\lag T^{ty}(-q)J^t(q)\rag_{-B}=-\frac{1}{2}iq\xi(B) B, \nonumber\\
  &\lag T^{ty}(-q)J^z_5(q)\rag_{-B}=-\frac{1}{2}iq\s(B).
\end{align}
Note that $J^t(q)$ and $J^z_5(q)$ couple to sources $V_t(-q)$ and $A_z(-q)$ respectively. We can rewrite \eqref{cross_response} in a more intuitive way
\begin{align}\label{Tty_response}
  &\lag T^{ty}(-q)\rag_{-B}=\frac{1}{2}\xi(B)BE_x(-q), \nonumber\\
  &\lag T^{ty}(-q)\rag_{-B}=\frac{1}{2}\s(B) B_{5y}(-q).
\end{align}

The thermal Hall effect contains both non-anomalous and anomalous contributions, where the non-anomalous contribution can be understood from non-anomalous MHD \cite{Hernandez:2017mch}. Naively, turning on $E_x$ necessarily induces steady flow along $v_y$ due to Lorentz force acting on positive and negative charge carriers. However, this is not true for a stationary state. 
The stationary state can be obtained simply by setting $\o=0$ in \eqref{cons_Ery} through \eqref{del_Az2}. In this case, the dynamics of the fields $\d g_{ty},\;\d V_t,\;\d A_z$ decouple from $\d V_x,\; \d g_{xy}$. We can thus consistently set $\d V_x$ and $\d g_{xy}$ to zero, leading to vanishing $J^x$ and $T^{xy}$. This strongly constrains the hydrodynamic analysis.
Note that both $J^x$ and $T^{xy}$ contain the dissipative terms as follows
\begin{align}
  &J^x=\s\left(E_x-\pd_x\m+v_yB\right),\nonumber\\
  &T^{xy}=\eta\pd_xv_y+\cdots\nonumber.
\end{align}
The shear contribution in $T^{xy}$ cannot be canceled by other terms. The only possibility for $T^{xy}$ to vanish is to have $v_y=0$. This implies an inhomogeneous vector charge density is needed for the stationary state: $E_x-\pd_x\m=0$. Indeed, this is consistent with the holographic analysis if we identify $\m=V_t(r=\infty)-V_t(r=r_h)$. 
With $v_y=0$, the non-anomalous contribution to thermal current is given by \cite{Hernandez:2017mch,Grozdanov:2016tdf} (see also \cite{Huang:2011dc,Finazzo:2016mhm}):
\begin{align}\label{Ex}
  T^{ty}=-E_xB\left(2p_{,B^2}-M_{\O,\m}\right).
\end{align}
The first term can be identified as $-E_xM$ by noting $2p_{,B^2}B=M$. The second term can be interpret as $-P_xB$ if we identify $E_xM_{\O,\m}$ as an effective polarization $P_x$. Apart from the non-anomalous contribution, we also obtain anomalous contribution that requires at least chiral anomaly to exist. This can be seen from the middle equation in \eqref{eom_q0}. When $\a=0$, the dynamics of $\d A_z$ decouples from that of $\d V_t$, leaving only non-anomalous contribution.
On the contrary, the thermal axial magnetic effect contains only anomalous effect. Its existence relies on gravitational anomaly.

\section{Conclusion} \label{conclusion}

In this work, based on a holographic model, we considered effects of the magneto-vortical coupling on transport properties of a strongly coupled plasma. First of all, even when the chiral and gravitational anomalies are turned off, the coupling of a magnetic field and a weak fluid's vorticity dynamically generates a contribution to the vector charge density $J^t$, which we refer to as non-anomalous medium contribution. 
The non-anomalous medium contribution in $J^t$ grows linearly in $B$ and the relevant transport $\xi$ approaches a constant in the strong magnetic field limit. However, similar non-anomalous contribution is not observed for the axial current.

Secondly, the magneto-vortical coupling also generates anomalous contributions to the vector charge density $J^t$ and axial current $J_5^z$. 
Thanks to the absence of a background for the vector chemical potential, the anomalous contribution to $J_5^z$ is completely induced by the gravitational anomaly (i.e., insensitive to the chiral anomaly). In this sense, it would be more natural to interpret the anomalous contribution to $J_5^z$ as CVE contribution \cite{1107.0368}, rather than CSE contribution. The corresponding chiral vortical conductivity receives correction at finite $B$. In particular, in the strong magnetic field limit, the magneto-vortical coupling renders the CVE conductivity to vanish asymptotically. This is quantitatively different from the conclusion of \cite{Hattori:2016njk}.
In contrast to that of $J_5^z$, the anomalous contribution to $J^t$ requires chiral anomaly to exist. The presence of gravitational anomaly can also affect the generation of $J^t$. Thus, the anomalous contribution to $J^t$ would contain more fruitful physics. Particularly, in the strong magnetic field limit, the anomalous part of $J^t$ seemingly grows logarithmically as a function of $B$. This is to be compared with the results of \cite{Hattori:2016njk} where a linear in $B$ term was generated to $J^t$ by the chiral anomaly.


Our findings summarized above would necessitate the formulating of a consistent/complete anomalous magnetohydrodynamics. This requires to consistently add novel transport phenomena induced by anomalies into the non-anomalous magnetohydrodynamics \cite{Hernandez:2017mch,Grozdanov:2016tdf}. While our study treated magentic field as external, the case with dynamical electromagnetic field is also interesting. The corresponding anomalous MHD has been initially considered in \cite{Hattori:2017usa} by assuming a small chiral anomaly coefficient. In fact, holographic models corresponding to dynamical electromagnetic field have been proposed in \cite{Grozdanov:2017kyl,Hofman:2017vwr}. Including anomalies to the model would allow us to study anomalous MHD without further assumption on the anomaly coefficients. We leave it for future work.

Last but not least, the transport phenomena we discussed are dissipationless. It is possible to derive them by including anomalies to the partition function approach \cite{Jensen:2012kj}. This would allow us to obtain more complete dissipationless transport phenomena. We hope to address this in the future.


\appendix

\section{Details of solving the background metric functions} \label{details_background}

In this appendix, we collect calculational details of solving the background metric functions.

When the magnetic field is weak (i.e. $B/T^2\ll1$), we construct the bulk metric functions perturbatively \cite{1202.2161}:
\begin{align}
f(r)&=r^2-\frac{r_h^4}{r^2}+\epsilon^2 f^{(2)}(r) +\epsilon^4 f^{(4)}(r)+\cdots, \nonumber \\
W_T(r)&=\log r+ \epsilon^2 W_T^{(2)}(r)+ \epsilon^4 W_T^{(4)}(r) +\cdots, \nonumber \\
W_L(r)&=\log r+ \epsilon^2 W_L^{(2)}(r)+ \epsilon^4 W_L^{(4)}(r) +\cdots,
\end{align}
where a formal parameter $\epsilon \sim B/T^2$  is introduced to mark the perturbative expansion.
$f^{(n)}(r)$, $W_T^{(n)}(r)$ and $W_L^{(n)}(r)$ are solved from the bulk equations. To our interest, we will be limited to $n=2$.

At $\mathcal{O}(\epsilon^2)$, the constraint equation \eqref{Wtl} yields,
\begin{align}
W_L^{(2)}+ 2 W_T^{(2)}= c_1+ \frac{c_2}{r},
\end{align}
where the integration constant $c_1$ should be set to zero due to asymptotic $AdS$ requirement \eqref{AdS_requirement}. By redefinition of the radial coordinate $r$, we could also set $c_2=0$. Thus,
\begin{align} \label{WT+L=0}
W_L^{(2)}+ 2 W_T^{(2)}=0.
\end{align}
Then, the dynamical equation \eqref{f} is solved as
\begin{align} \label{f2_pre}
&r^3\partial_r^2f^{(2)}+3r^2 \partial_r f^{(2)} = \frac{B^2}{3r} \Rightarrow f^{(2)}=c_f^1+ \frac{c_f^2}{r^2} - \frac{B^2\log r}{6r^2}.
\end{align}
Substituting \eqref{WT+L=0} and \eqref{f2_pre} into the dynamical equations \eqref{Wt} and \eqref{Wl}, we obtain
\begin{align}
&0=\left(r^5-r r_h^4\right) W_T^{(2)\prime\prime}(r)+\left(5 r^4-r_h^4\right) W_T^{(2)\prime}(r) +\frac{B^2}{6 r}+2 c_f^1 r, \nonumber \\
&0=\left(r^5-r r_h^4\right) W_L^{(2)\prime\prime}(r)+\left(5 r^4-r_h^4\right) W_L^{(2)\prime}(r) -\frac{B^2}{3 r}+2 c_f^1 r.
\end{align}
Obviously, in order to be consistent with \eqref{WT+L=0}, one has to set $c_f^1=0$. The integration constant $c_f^2$ is fixed by the location of the horizon, which will be presumably shifted due to the presence of a magnetic field,
\begin{align}
\left(r^2-\frac{r_h^4}{r^2}+ f^{(2)}(r)\right)\Bigg|_{r=r_p^{(2)}}=0 \Rightarrow c_f^2= -\left((r_p^{(2)})^4-r_h^4\right)+ \frac{B^2\log r_p^{(2)}}{6(r_p^{(2)})^2},
\end{align}
where $r_p^{(2)}$ represents the location of the event horizon at $\mathcal{O}(\epsilon^2)$. Then,
\begin{align} \label{f2_pert}
f^{(2)}= - \frac{(r_p^{(2)})^4-r_h^4}{r^2}+ \frac{B^2}{6r^2} \log\frac{r_p^{(2)}}{r}.
\end{align}
From \eqref{Wt}, $W_T^{(2)}$ is solved as
\begin{align}
&\partial_r\left[r(r^4-r_h^4)\partial_r W_T^{(2)}(r)\right]+\frac{B^2}{6 r}=0 \nonumber \\
&\Rightarrow W_T^{(2)}=\frac{B^2}{6}\int_r^\infty \frac{\log(x/r_h)}{x(x^4-r_h^4)}dx= \frac{B^2}{6}\int_r^\infty \frac{\log(x/r_p^{(2)})}{x(x^4-(r_p^{(2)})^4)}dx \nonumber \\
&=\frac{B^2}{288 r_h^4}
\left\{\pi^2 -24\log^2\frac{r_h}{r}\log\left(1-\frac{r_h^4}{r^4}\right)+ 12i\pi \log\frac{r}{r_h} -3 \text{Li}_2\left(\frac{r^4}{r_h^4}\right)\right\}, \label{WT2}
\end{align}
where in the last equality of the second line we have made use of the fact that the difference between $r_p^{(2)}$ and $r_h$ is of $\mathcal{O}(B^2)$. At $\mathcal{O}(\epsilon^2)$,  the relation between the location of the event horizon and the Hawking temperature becomes
\begin{align}
r_p^{(2)}-\frac{B^2}{24(r_p^{(2)})^3}+ \mathcal{O}(B^4)= \pi T,
\end{align}
which is solved as
\begin{align} \label{rp_perturb}
r_p^{(2)}=r_h+ \frac{B^2}{24 r_h^3} + \mathcal{O}(B^4),\qquad {\rm with} \qquad r_h=\pi T.
\end{align}

When the value of $B$ is generic, we have to solve the metric functions numerically. We find it more convenient to make a change of variables
\begin{align}
f(r) \to r^2 U(r),\qquad W_T(r)\to \log r + \frac{1}{2} \log V(r), \qquad W_L(r)\to \log r + \frac{1}{2} \log W(r),
\end{align}
followed by
\begin{align}
u=\frac{r_h}{r}\in [0,1].
\end{align}
Then, the dynamical bulk equations \eqref{f}-\eqref{Wl} turn into
\begin{align}
0=&U''(u) +U'(u) \left(\frac{V'(u)}{V(u)}+\frac{W'(u)}{2 W(u)}-\frac{5}{u}\right) +U(u) \left(\frac{8}{u^2}-\frac{2 V'(u)}{u V(u)}-\frac{W'(u)}{u W(u)}\right) \nonumber\\
&-\frac{B^2 u^2}{3 V(u)^2} -\frac{8}{u^2}, \nonumber\\
0=&V''(u)+ V'(u) \left(\frac{U'(u)}{U(u)}+\frac{W'(u)}{2 W(u)}-\frac{5}{u}\right) +V(u) \left(-\frac{8}{u^2 U(u)}+\frac{8}{u^2} -\frac{2 U'(u)}{u U(u)} \right. \nonumber\\
&\left.-\frac{W'(u)}{u W(u)}\right)+ \frac{2 B^2 u^2}{3 U(u) V(u)} , \nonumber\\
0=&W''(u)+ W'(u) \left(\frac{U'(u)}{U(u)}+\frac{V'(u)}{V(u)}-\frac{4}{u}\right) -\frac{W'(u)^2}{2 W(u)}+ W(u) \left(\frac{-\frac{B^2 u^2}{3 V(u)^2} -\frac{8}{u^2}} {U(u)}  \right. \nonumber\\
&\left. + \frac{8}{u^2}-\frac{2 U'(u)}{u U(u)}-\frac{2 V'(u)}{u V(u)}\right), \label{UVWeom}
\end{align}
where, since we have set $r_h=1$ above, $B$ should be understood as $B/r_h^2$.

Near the $AdS$ boundary $u=0$, the metric functions $U,V,W$ are expanded as:
\begin{align}
&U(u\to 0)= 1+ U_b^1u+ \frac{1}{4}(U_b^1)^2 u^2+ \frac{B^2}{6(V_b^0)^2} u^4\log u  + U_b^4 u^4 + \cdots, \nonumber \\
&V(u\to 0)= V_b^0 + V_b^0 U_b^1u+ \frac{1}{4}V_b^0(U_b^1)^2 u^2- \frac{B^2}{12 V_b^0} u^4\log u + V_b^4 u^4 + \cdots , \nonumber \\
&W(u\to 0)= W_b^0 + W_b^0 U_b^1 u + \frac{1}{4}W_b^0 (U_b^1)^2 u^2+ \frac{W_b^0 B^2}{6 (V_b^0)^2} u^4\log u + W_b^4 u^4 \cdots , \label{UVW_asymp}
\end{align}
where we have made use of the constraint equation \eqref{Wtl}. Obviously, the asymptotic boundary conditions only give rise to ``two'' effective requirements! The regularity requirements will yield another three conditions.
Just as in the fixing of $c_2$, we can utilise the freedom of redefining the radial coordinate $u$ and set $U_b^1=0$.

To summarise, the boundary conditions at $u=0$ (the $AdS$ boundary) are
\begin{align} \label{UVW_boundary}
U'(u=0)=0,\qquad V(u=0)=W(u=0) = 1,
\end{align}
while at the event horizon $u=1$
\begin{align}
&U(u=1)=0, \nonumber\\
&U'(1)V'(1)-8V(1)-2U'(1)V(1)+\frac{2B^2}{3V(1)}=0, \nonumber\\
&U'(1) W'(1)-2 W(1) U'(1)-8 W(1) -\frac{B^2 W(1)}{3 V(1)^2}=0. \label{UVW_horizon}
\end{align}

To find out the numeric solutions, one can proceed in two different ways. The first approach would be to directly solve \eqref{UVWeom} under the boundary conditions \eqref{UVW_boundary} and \eqref{UVW_horizon}. A second approach would be to replace the boundary conditions \eqref{UVW_boundary} by the following conditions at the horizon:
\begin{align}
&U(u\to 1)= 0+ U^1_h(u-1)+ U_h^2(u-1)^2+\cdots, \nonumber \\
&V(u\to 1)=V_h^0 + V_h^1(u-1)+\cdots, \nonumber \\
&W(u\to 1)= W_h^0+ W_h^1(u-1)+ \cdots,
\end{align}
where
\begin{align} \label{UVW_horizon1}
U_h^1=-4, \qquad V_h^0=1, \qquad W_h^0=1.
\end{align}
Note the choice of $U_h^1$ will set $\pi T=1$. However, solving \eqref{UVWeom} under the initial conditions \eqref{UVW_horizon} and \eqref{UVW_horizon1}, near the boundary $u=0$ the solution will behave as
\begin{align} \label{UVW_boundary1}
U(u\to 0) \to 1,\qquad V(u\to 0) \to v(b), \qquad W(u\to 0) \to w(b).
\end{align}
Then, the correct solution would be obtained by a further rescaling of the boundary coordinates
\begin{align} \label{xyz_rescale}
x\to \sqrt{v(b)}x, \qquad y\to \sqrt{v(b)}y,\qquad z\to \sqrt{w(b)}z.
\end{align}
Due to the ``incorrect'' asymptotic boundary behavior \eqref{UVW_boundary1}, we have relabeled the magnetic field by $b$ in \eqref{UVW_boundary1} and \eqref{xyz_rescale}. When solving the EOMs \eqref{UVWeom} under the initial conditions \eqref{UVW_horizon} and \eqref{UVW_horizon1}, the same relabeling should be made. Recalling the definition of the magnetic field $F^V= bdx \wedge dy$, the physical magnetic field $B$ (in unit of $r_h^2$) should be
\begin{align}
B= \frac{b}{v(b)}.
\end{align}
Finally, we would like to point out that the background solution obtained with conditions \eqref{UVW_horizon} and \eqref{UVW_horizon1} does not necessarily satisfy $U_b^1=0$ (cf. \eqref{UVW_asymp}).

\section{Horizon boundary conditions from matching} \label{horizon_matching}

We first seek solutions to \eqref{cons_Ery} through \eqref{del_Az2} near the horizon with ingoing boundary conditions. We obtain the following series solutions
\begin{align}\label{h_sol}
  &\d A_z=a_0(r-r_h)^\b+a_1(r-r_h)^{\b+1}+\cdots,\nonumber\\
  &\d V_{x}=b_0(r-r_h)^\b+b_1(r-r_h)^{\b+1}+\cdots,\nonumber\\
  &\d g_{ty}=c_1(r-r_h)^{\b+1}+c_2(r-r_h)^{\b+2}+\cdots,\nonumber\\
  &\d V_t=d_1(r-r_h)^{\b+1}+d_2(r-r_h)^{\b+2}+\cdots,\nonumber\\
  &\d g_{xy}=e_0(r-r_h)^\b+e_1(r-r_h)^{\b+1}+\cdots,
\end{align}
with $\b=-\frac{i\o}{f'(r_h)}$. Here, $a_0$, $b_0$ and $e_0$ are free parameters, while all the rest coefficients are completely determined by them. For instance, $c_1$ and $d_1$ are
\begin{align}\label{recursive}
  &c_1=\frac{if'(r_h)}{f'(r_h)-i\o}\left\{-iBe^{-4W_T(r_h)} b_0 -4\l B^2 e^{-W_L(r_h)-6W_T(r_h)} q a_0 \right.  \nonumber\\
  & \left. \qquad +e^{-W_L(r_h)-2W_T(r_h)}q\left[e^{W_L(r_h)}e_0+4\l f'(r_h) \left(W_L'(r_h) +2W_T'(r_h)\right)a_0 \right]\right\},\nonumber\\
  &d_1=\frac{if'(r_h)}{f'(r_h)-i\o} e^{-W_L(r_h)-2W_T(r_h)} \left[e^{W_L(r_h)} q b_0 + 8iB\a a_0\right].
\end{align}
The three parameters $a_0$, $b_0$ and $e_0$ do not match the five sources to the fields $\d A_z$, $\d V_x$, $\d g_{ty}$, $\d V_t$ and $\d g_{xy}$. The remaining two parameters come from pure gauge solutions, which are gauge transformation of trivial solution:
\begin{align}\label{pure}
  &\d V_x=-BC_1,\quad \d g_{ty}=-i\o C_1,\quad \d g_{xy}=iqC_1,\quad \text{others}=0;\nonumber\\
  &\d V_t=-i\o C_2,\quad \d V_x=iqC_2,\quad \text{others}=0.
\end{align}
The horizon solutions are to be matched with the lowest order solutions in \eqref{omega_exp} near the horizon region. Since the horizon solutions also contain $\d V_x$ and $\d g_{xy}$, we also need the lowest order solutions to them. To the lowest order in $\o$, the EOMs of $\d V_x$ and $\d g_{xy}$ decouple:
\begin{align}
  &\frac{1}{2}iq\pd_r\d g_{xy}+\frac{1}{2}Be^{-2W_T}\pd_r\d V_x=0,\nonumber\\
  &\pd_r\left(e^{W_L+2W_T}f\pd_r\d g_{xy}\right)=0,\nonumber\\
  &\pd_r\left(e^{W_L}f\pd_r\d V_{x}\right)=0.
\end{align}
They are clearly solved by constant solutions. Matching with the horizon solutions, we simply have
\begin{align}
  \d V_x=b_0,\quad \d g_{xy}=e_0.
\end{align}
Note that we can set the above two solutions to zero by adding pure gauge solutions. In the limit $\o\to0$, the pure gauge solutions do not change the horizon values of $\d g_{ty}$, $\d V_t$ and $\d A_z$:
\begin{align}
  \d g_{ty}(r=r_h)=0,\quad \d V_t(r=r_h)=0,\quad \d A_z(r=r_h)={\rm constant}.
\end{align}
The $\o\to0$ limit of \eqref{recursive} determines the horizon derivative of $\d g_{ty}$ and $\d V_t$. For the decoupled EOMs \eqref{eom_gty} through \eqref{eom_az}, we can take the horizon derivatives of $\d g_{ty}$ and $\d V_t$, and horizon value of $\d A_z$ as free parameters.

\section{Magnetic and magneto-vortical susceptibilities}\label{app_suscept}

In this appendix, we calculate the magnetic susceptibility $2p_{,B^2}$ and magneto-vortical susceptibility $M_{\O}$ independently as a confirmation to our claim in \eqref{suscept}.

Let us begin with the magnetic susceptibility $2p_{,B^2}$ and the magnetization $M$.
For the equilibrium state (corresponding to the magnetic brane background), the stress tensor for the boundary theory is computed as
\begin{align}
&T^{tt}=\lim_{r\to\infty} \left\{-2r^6 \left[-\frac{3}{f(r)} +\frac{W_L^\prime + 2 W_T^\prime}{\sqrt{f(r)}} - \frac{B^2 e^{-4W_T}r^6 \log r}{2f(r)}  \right]\right\}, \nonumber \\
&T^{xx}=T^{yy}=\lim_{r\to \infty}\left\{-2r^6 \left[3e^{-2W_T}- \frac{e^{-2W_T}f^\prime(r)}{2\sqrt{f(r)}} - \sqrt{f(r)} e^{-2 W_T} (W_L^\prime+ W_T^\prime)  \right] - \frac{1}{2}B^2 e^{-6 W_T} r^6 \log r \right\}, \nonumber \\
&T^{zz}=\lim_{r\to\infty} \left\{- 2r^6 \left[3e^{-2W_L}- \frac{e^{-2W_L}f^\prime(r)} {2\sqrt{f(r)}} -2 e^{-2W_L} \sqrt{f(r)} W_T^\prime\right]+ \frac{1}{2} B^2 e^{-2W_L- 4W_T} r^6\log r\right\}.
\end{align}
With the analytical solution presented in appendix \ref{details_background}, it is straightforward to compute the various components of $T^{\mu\nu}$:
\begin{align} \label{Tmunu_pert}
&T^{tt}=3r_h^4+3(r_p^4-r_h^4)- \frac{1}{2}B^2\log r_h + \mathcal{O}(B^3), \nonumber \\
&T^{xx}=T^{yy}= r_h^4+(r_p^4-r_h^4) -\frac{1}{6}B^2 - \frac{1}{2}B^2 \log r_h + \mathcal{O}(B^3), \nonumber \\
&T^{zz}= r_h^4+(r_p^4-r_h^4) - \frac{1}{6}B^2+ \frac{1}{2}B^2 \log r_h + \mathcal{O}(B^3).
\end{align}
To extract the energy density, pressure and magnetization, we compare \eqref{Tmunu_pert} with the MHD formalism \cite{Huang:2011dc} (see equation (14) there). Here, we would like to point out that the AdS/CFT computations give rise to the medium contributions (denoted as $T^{\mu\nu}_{\rm F0}$ in \cite{Huang:2011dc}). Consequently,
\begin{align}
\varepsilon= T^{tt}, \qquad p_\perp= T^{xx}= T^{yy}, \qquad p_\parallel= T^{zz},
\end{align}
The pressure $p$ is identified as $p_\parallel$
\begin{align}
p=p_\parallel= \frac{1}{2}B^2 \log r_h + \mathcal{O}(B^3)\Longrightarrow 2p_{,B^2}= \log r_h,
\end{align}
where we used the perturbative expression for $r_p$ in \eqref{rp_perturb}.
The magnetization $M$ could be computed as
\begin{align}
p_\perp= p_\parallel-MB \Longrightarrow M= B \log r_h+ \mathcal{O}(B^2).
\end{align}

Now we move on to the calculation of the magneto-vortical susceptibility $M_\O$. In the zero magnetic field situation, we calculate $M_\O$ based on the following Kubo formula \cite{Hernandez:2017mch}:
\begin{align} \label{kubo_M_omega}
M_\Omega= -\lim_{q_x, q_y\to 0} \frac{\langle T^{ty} J^x \rangle}{q_y q_x}.
\end{align}
Since $M_\O$ is C-odd, we need to consider the finite density RN-$AdS_5$ background:
\begin{align} \label{RN_AdS5}
&ds^2= -f(r) dt^2+ \frac{dr^2}{f(r)} + r^2(dx^2+ dy^2+ dz^2), \nonumber\\
&V= \left(\frac{Q}{r_h^2}- \frac{Q}{r^2}\right)dt, \qquad A=0,
\end{align}
where
\begin{align}
f(r)= r^2\left(1-\frac{r_h^4}{r^4} + \frac{Q^2}{3r^6}- \frac{Q^2}{3r_h^2 r^4}\right).
\end{align}

For consistency, we turn on the following fluctuations on top of \eqref{RN_AdS5},
\begin{align} \label{fluc}
&\delta (ds^2)= 2r^2 \left[\delta g_{tx}(r,x,y)dtdx+ \delta g_{ty}(r,x,y)dtdy \right], \nonumber \\
&\delta V= \delta V_x(r,x,y)dx+ \delta V_y(r,x,y)dy.
\end{align}
We turn to the Fourier space by assuming plane wave ansatz for the fluctuations,
\begin{align}
&\delta g_{tx}(r,x,y)\sim e^{i q_x x+ iq_y y} \delta g_{tx}(r), \qquad \qquad \delta g_{ty}(r,x,y)\sim e^{i q_x x+ iq_y y} \delta g_{ty}(r),\nonumber \\
&\delta V_x(r,x,y)\sim e^{i q_x x+ iq_y y} \delta V_x(r), \qquad \qquad \delta V_y(r,x,y)\sim e^{i q_x x+ iq_y y} \delta V_y(r).
\end{align}
To  proceed, we collect the EOMs for the fluctuations in \eqref{fluc}. The constraint equations $E_{rt}=0$ and $EV^r=0$ give rise to
\begin{align}
&r^{-2}f(r)\partial_r\left(q_x \delta g_{tx} + q_y \delta g_{ty}\right)- \left(r^{-2}f(r) \right)^\prime\left(q_x \delta g_{tx} + q_y \delta g_{ty}\right)=0, \\
&rf(r) \partial_r(q_x \delta V_x+ q_y \delta V_y) + 2Q (q_x \delta g_{tx} + q_y \delta g_{ty})=0.
\end{align}
The dynamical components of the Einstein equations $E_{tx}=E_{ty}=0$ are
\begin{align}
&\partial_r(r^5 \partial_r \delta g_{tx}) + 2Q \partial_r \delta V_x +\frac{r^3}{f(r)} \left(q_x q_y \delta g_{ty}- q_y^2 \delta g_{tx}\right)=0, \label{gtx}\\
&\partial_r(r^5 \partial_r \delta g_{ty}) + 2Q \partial_r \delta V_y +\frac{r^3}{f(r)} \left(q_x q_y \delta g_{tx}- q_x^2 \delta g_{ty}\right)=0. \label{gty}
\end{align}
The dynamical components of the vector Maxwell equations $EV^x=EV^y=0$ are
\begin{align}
&\partial_r\left[rf(r)\partial_r \delta V_x\right] + 2Q \partial_r \delta g_{tx} + \frac{1}{r} \left(q_x q_y \delta V_y- q_y^2 \delta V_x\right)=0, \label{Vx}\\
&\partial_r\left[rf(r)\partial_r \delta V_y\right] + 2Q \partial_r \delta g_{ty} + \frac{1}{r} \left(q_x q_y \delta V_x- q_x^2 \delta V_y\right)=0. \label{Vy}
\end{align}

Near the AdS boundary, we impose
\begin{align}
\delta V_x \xrightarrow[]{r\to \infty} v_x, \qquad {\rm others} \xrightarrow[]{r\to \infty} 0,
\end{align}
while at the horizon we have
\begin{align}
\delta g_{tx}, \delta g_{ty} \xrightarrow[]{r\to r_h} 0, \qquad \delta V_x, \delta V_y ~{\rm are~regualr~at}~r=r_h.
\end{align}
Given \eqref{kubo_M_omega}, we solve \eqref{gtx}-\eqref{Vy} in the small momenta limit. The fluctuation modes could be expanded as
\begin{align}
&\delta g_{tx}= \delta g_{tx}^{(0)} + \epsilon \,\delta g_{tx}^{(1)} + \epsilon^2 \delta g_{tx}^{(2)}, \nonumber\\
&\delta g_{ty}= \delta g_{ty}^{(0)} + \epsilon \,\delta g_{ty}^{(1)} + \epsilon^2 \delta g_{ty}^{(2)}, \nonumber\\
&\delta V_x= \delta V_x^{(0)} + \epsilon \,\delta V_x^{(1)} + \epsilon^2 \delta V_x^{(2)}, \nonumber\\
&\delta V_y= \delta V_y^{(0)} + \epsilon \, \delta V_y^{(1)} + \epsilon^2 \delta V_y^{(2)},
\end{align}
where $\epsilon \sim q_x, q_y$.

At the lowest order $\mathcal{O}(\epsilon^0)$, the solutions are simply given by
\begin{align}
\delta V_x^{(0)}= v_x, \qquad \delta g_{tx}^{(0)}= \delta g_{ty}^{(0)}= \delta V_y^{(0)}=0.
\end{align}
At the first order $\mathcal{O}(\epsilon^1)$, there are no non-trivial solutions
\begin{align}
\delta g_{tx}^{(1)}= \delta g_{tx}^{(1)}= \delta V_y^{(1)}= \delta V_y^{(1)}=0.
\end{align}
At the second order $\mathcal{O}(\epsilon^2)$, the equations we need are
\begin{align}
&\partial_r(r^5 \partial_r \delta g_{ty}^{(2)}) +2 Q \partial_r \delta V_y^{(2)}=0, \label{gty_2nd}\\
&\partial_r\left[rf(r)\partial_r \delta V_y^{(2)}\right] + 2Q \partial_r \delta g_{ty}^{(2)} + \frac{q_x q_y}{r} \delta V_x^{(0)}=0. \label{Vy_2nd}
\end{align}
In $\delta g_{ty}^{(2)}$, we will track the term linear in $Q$ ($\sim \mu$) only. Therefore, the $Q$-term in \eqref{Vy_2nd} could be discarded. Furthermore, we can simply take $f(r) \to r^2(1-r_h^4/r^4)$. The equation \eqref{Vy_2nd} is solved as
\begin{align}
\delta V_y^{(2)}= -q_x q_y v_x \int_{r}^{\infty} \frac{\log(r_h/x)} {x^3(1-r_h^4/x^4)} dx.
\end{align}
Finally, the equation \eqref{gty_2nd} is solved as
\begin{align}
\delta g_{ty}^{(2)}= 2Q q_x q_y v_x \int_r^\infty \frac{dx}{x^5} \int_{r_h}^x  \frac{\log (r_h/y)}{y^3(1-r_h^4/y^4)}dy + \frac{C_0}{r^4},
\end{align}
where the integration constant $C_0$ is fixed as
\begin{align}
2Q q_x q_y v_x \int_{r_h}^\infty \frac{dx}{x^5} \int_{r_h}^x  \frac{\log (r_h/y)}{y^3(1-r_h^4/y^4)}dy + \frac{C_0}{r_h^4}=0 \Rightarrow C_0= \frac{\pi^2-8}{64r_h^2} Q q_x q_y v_x.
\end{align}
Near the $AdS$ boundary,
\begin{align}
\delta g_{ty}^{(2)} \xrightarrow[]{r\to \infty} - \frac{Qq_x q_y v_x}{8 r_h^2 r^4}+ \mathcal{O}(r^{-5}),
\end{align}
which is translated to 
\begin{align}
T^{ty}= \frac{Q}{2r_h^2} q_x q_y v_x \Rightarrow M_\Omega= -\frac{Q}{2r_h^2} = - \frac{1}{2} \mu.
\end{align}

\section*{Acknowledgements}

We would like to thank K. Hattori, J.F. Liao, K. Mameda, R.X. Miao, I. Shovkovy, H.-U. Yee and Y. Yin for useful discussions related to this work.
YB was supported by the Fundamental Research Funds for the Central Universities under grant No.122050205032 and the Natural Science Foundation of China (NSFC) under the grant No.11705037. SL was supported by the NSFC under the grant Nos. 11675274 and 11735007. SL also thanks Yukawa Institute of Theoretical Physics for hospitality and the workshop ``Quantum kinetic theories in magnetic and vortical fields'' for providing a stimulating environment during the final stage of this work.

\bibliographystyle{utphys}
\bibliography{reference}

\providecommand{\href}[2]{#2}\begingroup\raggedright\begin{thebibliography}{10}

\bibitem{Bali:2012zg}
G.~S. Bali, F.~Bruckmann, G.~Endrodi, Z.~Fodor, S.~D. Katz, and A.~Schafer,
  ``{QCD quark condensate in external magnetic fields},''
  \href{http://dx.doi.org/10.1103/PhysRevD.86.071502}{{\em Phys. Rev.}
  {\bfseries D86} (2012) 071502},
\href{http://arxiv.org/abs/1206.4205}{{\ttfamily arXiv:1206.4205 [hep-lat]}}.

\bibitem{Bali:2011qj}
G.~S. Bali, F.~Bruckmann, G.~Endrodi, Z.~Fodor, S.~D. Katz, S.~Krieg,
  A.~Schafer, and K.~K. Szabo, ``{The QCD phase diagram for external magnetic
  fields},'' \href{http://dx.doi.org/10.1007/JHEP02(2012)044}{{\em JHEP}
  {\bfseries 02} (2012) 044},
\href{http://arxiv.org/abs/1111.4956}{{\ttfamily arXiv:1111.4956 [hep-lat]}}.

\bibitem{Chernodub:2011mc}
M.~N. Chernodub, ``{Spontaneous electromagnetic superconductivity of vacuum in
  strong magnetic field: evidence from the Nambu--Jona-Lasinio model},''
  \href{http://dx.doi.org/10.1103/PhysRevLett.106.142003}{{\em Phys. Rev.
  Lett.} {\bfseries 106} (2011) 142003},
\href{http://arxiv.org/abs/1101.0117}{{\ttfamily arXiv:1101.0117 [hep-ph]}}.

\bibitem{Chernodub:2010qx}
M.~N. Chernodub, ``{Superconductivity of QCD vacuum in strong magnetic
  field},'' \href{http://dx.doi.org/10.1103/PhysRevD.82.085011}{{\em Phys.
  Rev.} {\bfseries D82} (2010) 085011},
\href{http://arxiv.org/abs/1008.1055}{{\ttfamily arXiv:1008.1055 [hep-ph]}}.

\bibitem{Jiang:2016wvv}
Y.~Jiang and J.~Liao, ``{Pairing Phase Transitions of Matter under Rotation},''
  \href{http://dx.doi.org/10.1103/PhysRevLett.117.192302}{{\em Phys. Rev.
  Lett.} {\bfseries 117} no.~19, (2016) 192302},
\href{http://arxiv.org/abs/1606.03808}{{\ttfamily arXiv:1606.03808 [hep-ph]}}.

\bibitem{Ebihara:2016fwa}
S.~Ebihara, K.~Fukushima, and K.~Mameda, ``{Boundary effects and gapped
  dispersion in rotating fermionic matter},''
  \href{http://dx.doi.org/10.1016/j.physletb.2016.11.010}{{\em Phys. Lett.}
  {\bfseries B764} (2017) 94--99},
\href{http://arxiv.org/abs/1608.00336}{{\ttfamily arXiv:1608.00336 [hep-ph]}}.

\bibitem{Vilenkin:1980fu}
A.~Vilenkin, ``{EQUILIBRIUM PARITY VIOLATING CURRENT IN A MAGNETIC FIELD},''
\href{http://dx.doi.org/10.1103/PhysRevD.22.3080}{{\em Phys. Rev.} {\bfseries
  D22} (1980) 3080--3084}.

\bibitem{Fukushima:2008xe}
K.~Fukushima, D.~E. Kharzeev, and H.~J. Warringa, ``{The Chiral Magnetic
  Effect},'' \href{http://dx.doi.org/10.1103/PhysRevD.78.074033}{{\em Phys.
  Rev.} {\bfseries D78} (2008) 074033},
\href{http://arxiv.org/abs/0808.3382}{{\ttfamily arXiv:0808.3382 [hep-ph]}}.

\bibitem{Fukushima:2010vw}
K.~Fukushima, D.~E. Kharzeev, and H.~J. Warringa, ``{Real-time dynamics of the
  Chiral Magnetic Effect},''
  \href{http://dx.doi.org/10.1103/PhysRevLett.104.212001}{{\em Phys. Rev.
  Lett.} {\bfseries 104} (2010) 212001},
\href{http://arxiv.org/abs/1002.2495}{{\ttfamily arXiv:1002.2495 [hep-ph]}}.

\bibitem{Erdmenger:2008rm}
J.~Erdmenger, M.~Haack, M.~Kaminski, and A.~Yarom, ``{Fluid dynamics of
  R-charged black holes},''
  \href{http://dx.doi.org/10.1088/1126-6708/2009/01/055}{{\em JHEP} {\bfseries
  01} (2009) 055},
\href{http://arxiv.org/abs/0809.2488}{{\ttfamily arXiv:0809.2488 [hep-th]}}.

\bibitem{Banerjee:2008th}
N.~Banerjee, J.~Bhattacharya, S.~Bhattacharyya, S.~Dutta, R.~Loganayagam, and
  P.~Surowka, ``{Hydrodynamics from charged black branes},''
  \href{http://dx.doi.org/10.1007/JHEP01(2011)094}{{\em JHEP} {\bfseries 01}
  (2011) 094},
\href{http://arxiv.org/abs/0809.2596}{{\ttfamily arXiv:0809.2596 [hep-th]}}.

\bibitem{Son:2009tf}
D.~T. Son and P.~Surowka, ``{Hydrodynamics with Triangle Anomalies},''
  \href{http://dx.doi.org/10.1103/PhysRevLett.103.191601}{{\em Phys. Rev.
  Lett.} {\bfseries 103} (2009) 191601},
\href{http://arxiv.org/abs/0906.5044}{{\ttfamily arXiv:0906.5044 [hep-th]}}.

\bibitem{Metlitski:2005pr}
M.~A. Metlitski and A.~R. Zhitnitsky, ``{Anomalous axion interactions and
  topological currents in dense matter},''
  \href{http://dx.doi.org/10.1103/PhysRevD.72.045011}{{\em Phys. Rev.}
  {\bfseries D72} (2005) 045011},
\href{http://arxiv.org/abs/hep-ph/0505072}{{\ttfamily arXiv:hep-ph/0505072
  [hep-ph]}}.

\bibitem{Son:2004tq}
D.~T. Son and A.~R. Zhitnitsky, ``{Quantum anomalies in dense matter},''
  \href{http://dx.doi.org/10.1103/PhysRevD.70.074018}{{\em Phys. Rev.}
  {\bfseries D70} (2004) 074018},
\href{http://arxiv.org/abs/hep-ph/0405216}{{\ttfamily arXiv:hep-ph/0405216
  [hep-ph]}}.

\bibitem{Hattori:2016njk}
K.~Hattori and Y.~Yin, ``{Charge redistribution from anomalous magnetovorticity
  coupling},'' \href{http://dx.doi.org/10.1103/PhysRevLett.117.152002}{{\em
  Phys. Rev. Lett.} {\bfseries 117} no.~15, (2016) 152002},
\href{http://arxiv.org/abs/1607.01513}{{\ttfamily arXiv:1607.01513 [hep-th]}}.

\bibitem{Liu:2017spl}
Y.~Liu and I.~Zahed, ``{Pion Condensation by Rotation in a Magnetic field},''
  \href{http://dx.doi.org/10.1103/PhysRevLett.120.032001}{{\em Phys. Rev.
  Lett.} {\bfseries 120} no.~3, (2018) 032001},
\href{http://arxiv.org/abs/1711.08354}{{\ttfamily arXiv:1711.08354 [hep-ph]}}.

\bibitem{Chen:2015hfc}
H.-L. Chen, K.~Fukushima, X.-G. Huang, and K.~Mameda, ``{Analogy between
  rotation and density for Dirac fermions in a magnetic field},''
  \href{http://dx.doi.org/10.1103/PhysRevD.93.104052}{{\em Phys. Rev.}
  {\bfseries D93} no.~10, (2016) 104052},
\href{http://arxiv.org/abs/1512.08974}{{\ttfamily arXiv:1512.08974 [hep-ph]}}.

\bibitem{Cao:2019ctl}
G.~Cao and L.~He, ``{Rotation induced charged pion condensation in a strong
  magnetic field: A Nambu–Jona-Lasino model study},''
  \href{http://dx.doi.org/10.1103/PhysRevD.100.094015}{{\em Phys. Rev.}
  {\bfseries D100} no.~9, (2019) 094015},
\href{http://arxiv.org/abs/1910.02728}{{\ttfamily arXiv:1910.02728 [nucl-th]}}.

\bibitem{Chen:2019tcp}
H.-L. Chen, X.-G. Huang, and K.~Mameda, ``{Do charged pions condense in a
  magnetic field with rotation?},''
\href{http://arxiv.org/abs/1910.02700}{{\ttfamily arXiv:1910.02700 [nucl-th]}}.

\bibitem{Kovtun:2016lfw}
P.~Kovtun, ``{Thermodynamics of polarized relativistic matter},''
  \href{http://dx.doi.org/10.1007/JHEP07(2016)028}{{\em JHEP} {\bfseries 07}
  (2016) 028},
\href{http://arxiv.org/abs/1606.01226}{{\ttfamily arXiv:1606.01226 [hep-th]}}.

\bibitem{Neiman:2010zi}
Y.~Neiman and Y.~Oz, ``{Relativistic Hydrodynamics with General Anomalous
  Charges},'' \href{http://dx.doi.org/10.1007/JHEP03(2011)023}{{\em JHEP}
  {\bfseries 03} (2011) 023},
\href{http://arxiv.org/abs/1011.5107}{{\ttfamily arXiv:1011.5107 [hep-th]}}.

\bibitem{Landsteiner:2011cp}
K.~Landsteiner, E.~Megias, and F.~Pena-Benitez, ``{Gravitational Anomaly and
  Transport},'' \href{http://dx.doi.org/10.1103/PhysRevLett.107.021601}{{\em
  Phys. Rev. Lett.} {\bfseries 107} (2011) 021601},
\href{http://arxiv.org/abs/1103.5006}{{\ttfamily arXiv:1103.5006 [hep-ph]}}.

\bibitem{Landsteiner:2016led}
K.~Landsteiner, ``{Notes on Anomaly Induced Transport},''
  \href{http://dx.doi.org/10.5506/APhysPolB.47.2617}{{\em Acta Phys. Polon.}
  {\bfseries B47} (2016) 2617},
\href{http://arxiv.org/abs/1610.04413}{{\ttfamily arXiv:1610.04413 [hep-th]}}.

\bibitem{1006.2400}
T.~E. Clark, S.~T. Love, and T.~ter Veldhuis, ``{Holographic Currents and
  Chern-Simons Terms},''
  \href{http://dx.doi.org/10.1103/PhysRevD.82.106004}{{\em Phys. Rev.}
  {\bfseries D82} (2010) 106004},
\href{http://arxiv.org/abs/1006.2400}{{\ttfamily arXiv:1006.2400 [hep-th]}}.

\bibitem{1107.0368}
K.~Landsteiner, E.~Megias, L.~Melgar, and F.~Pena-Benitez, ``{Holographic
  Gravitational Anomaly and Chiral Vortical Effect},''
  \href{http://dx.doi.org/10.1007/JHEP09(2011)121}{{\em JHEP} {\bfseries 09}
  (2011) 121},
\href{http://arxiv.org/abs/1107.0368}{{\ttfamily arXiv:1107.0368 [hep-th]}}.

\bibitem{deHaro:2000vlm}
S.~de~Haro, S.~N. Solodukhin, and K.~Skenderis, ``{Holographic reconstruction
  of space-time and renormalization in the AdS / CFT correspondence},''
  \href{http://dx.doi.org/10.1007/s002200100381}{{\em Commun. Math. Phys.}
  {\bfseries 217} (2001) 595--622},
\href{http://arxiv.org/abs/hep-th/0002230}{{\ttfamily arXiv:hep-th/0002230
  [hep-th]}}.

\bibitem{Sahoo:2010sp}
B.~Sahoo and H.-U. Yee, ``{Electrified plasma in AdS/CFT correspondence},''
  \href{http://dx.doi.org/10.1007/JHEP11(2010)095}{{\em JHEP} {\bfseries 11}
  (2010) 095},
\href{http://arxiv.org/abs/1004.3541}{{\ttfamily arXiv:1004.3541 [hep-th]}}.

\bibitem{1304.5529}
E.~Megias and F.~Pena-Benitez, ``{Holographic Gravitational Anomaly in First
  and Second Order Hydrodynamics},''
  \href{http://dx.doi.org/10.1007/JHEP05(2013)115}{{\em JHEP} {\bfseries 05}
  (2013) 115},
\href{http://arxiv.org/abs/1304.5529}{{\ttfamily arXiv:1304.5529 [hep-th]}}.

\bibitem{1706.05294}
C.~Copetti, J.~Fernandez-Pendas, K.~Landsteiner, and E.~Megias, ``{Anomalous
  transport and holographic momentum relaxation},''
  \href{http://dx.doi.org/10.1007/JHEP09(2017)004}{{\em JHEP} {\bfseries 09}
  (2017) 004},
\href{http://arxiv.org/abs/1706.05294}{{\ttfamily arXiv:1706.05294 [hep-th]}}.

\bibitem{0908.3875}
E.~D'Hoker and P.~Kraus, ``{Magnetic Brane Solutions in AdS},''
  \href{http://dx.doi.org/10.1088/1126-6708/2009/10/088}{{\em JHEP} {\bfseries
  10} (2009) 088},
\href{http://arxiv.org/abs/0908.3875}{{\ttfamily arXiv:0908.3875 [hep-th]}}.

\bibitem{Hernandez:2017mch}
J.~Hernandez and P.~Kovtun, ``{Relativistic magnetohydrodynamics},''
  \href{http://dx.doi.org/10.1007/JHEP05(2017)001}{{\em JHEP} {\bfseries 05}
  (2017) 001},
\href{http://arxiv.org/abs/1703.08757}{{\ttfamily arXiv:1703.08757 [hep-th]}}.

\bibitem{Fuini:2015hba}
J.~F. Fuini and L.~G. Yaffe, ``{Far-from-equilibrium dynamics of a strongly
  coupled non-Abelian plasma with non-zero charge density or external magnetic
  field},'' \href{http://dx.doi.org/10.1007/JHEP07(2015)116}{{\em JHEP}
  {\bfseries 07} (2015) 116},
\href{http://arxiv.org/abs/1503.07148}{{\ttfamily arXiv:1503.07148 [hep-th]}}.

\bibitem{Gynther:2010ed}
A.~Gynther, K.~Landsteiner, F.~Pena-Benitez, and A.~Rebhan, ``{Holographic
  Anomalous Conductivities and the Chiral Magnetic Effect},''
  \href{http://dx.doi.org/10.1007/JHEP02(2011)110}{{\em JHEP} {\bfseries 02}
  (2011) 110},
\href{http://arxiv.org/abs/1005.2587}{{\ttfamily arXiv:1005.2587 [hep-th]}}.

\bibitem{Tatsumi:2014wka}
T.~Tatsumi, K.~Nishiyama, and S.~Karasawa, ``{Novel Lifshitz point for chiral
  transition in the magnetic field},''
  \href{http://dx.doi.org/10.1016/j.physletb.2015.02.033}{{\em Phys. Lett.}
  {\bfseries B743} (2015) 66--70},
\href{http://arxiv.org/abs/1405.2155}{{\ttfamily arXiv:1405.2155 [hep-ph]}}.

\bibitem{Bu:2018trt}
Y.~Bu and S.~Lin, ``{Holographic magnetized chiral density wave},''
  \href{http://dx.doi.org/10.1088/1674-1137/42/11/114104}{{\em Chin. Phys.}
  {\bfseries C42} no.~11, (2018) 114104},
\href{http://arxiv.org/abs/1807.00330}{{\ttfamily arXiv:1807.00330 [hep-th]}}.

\bibitem{Lin:2018aon}
S.~Lin and L.~Yang, ``{Mass correction to chiral vortical effect and chiral
  separation effect},''
  \href{http://dx.doi.org/10.1103/PhysRevD.98.114022}{{\em Phys. Rev.}
  {\bfseries D98} no.~11, (2018) 114022},
\href{http://arxiv.org/abs/1810.02979}{{\ttfamily arXiv:1810.02979 [nucl-th]}}.

\bibitem{Ji:2019pxx}
X.~Ji, Y.~Liu, and X.-M. Wu, ``{Chiral vortical conductivity across a
  topological phase transition from holography},''
  \href{http://dx.doi.org/10.1103/PhysRevD.100.126013}{{\em Phys. Rev.}
  {\bfseries D100} no.~12, (2019) 126013},
\href{http://arxiv.org/abs/1904.08058}{{\ttfamily arXiv:1904.08058 [hep-th]}}.

\bibitem{Flachi:2017vlp}
A.~Flachi and K.~Fukushima, ``{Chiral vortical effect with finite rotation,
  temperature, and curvature},''
  \href{http://dx.doi.org/10.1103/PhysRevD.98.096011}{{\em Phys. Rev.}
  {\bfseries D98} no.~9, (2018) 096011},
\href{http://arxiv.org/abs/1702.04753}{{\ttfamily arXiv:1702.04753 [hep-th]}}.

\bibitem{Kovtun:2012rj}
P.~Kovtun, ``{Lectures on hydrodynamic fluctuations in relativistic
  theories},'' \href{http://dx.doi.org/10.1088/1751-8113/45/47/473001}{{\em J.
  Phys.} {\bfseries A45} (2012) 473001},
\href{http://arxiv.org/abs/1205.5040}{{\ttfamily arXiv:1205.5040 [hep-th]}}.

\bibitem{Grozdanov:2016tdf}
S.~Grozdanov, D.~M. Hofman, and N.~Iqbal, ``{Generalized global symmetries and
  dissipative magnetohydrodynamics},''
  \href{http://dx.doi.org/10.1103/PhysRevD.95.096003}{{\em Phys. Rev.}
  {\bfseries D95} no.~9, (2017) 096003},
\href{http://arxiv.org/abs/1610.07392}{{\ttfamily arXiv:1610.07392 [hep-th]}}.

\bibitem{Huang:2011dc}
X.-G. Huang, A.~Sedrakian, and D.~H. Rischke, ``{Kubo formulae for relativistic
  fluids in strong magnetic fields},''
  \href{http://dx.doi.org/10.1016/j.aop.2011.08.001}{{\em Annals Phys.}
  {\bfseries 326} (2011) 3075--3094},
\href{http://arxiv.org/abs/1108.0602}{{\ttfamily arXiv:1108.0602
  [astro-ph.HE]}}.

\bibitem{Finazzo:2016mhm}
S.~I. Finazzo, R.~Critelli, R.~Rougemont, and J.~Noronha, ``{Momentum transport
  in strongly coupled anisotropic plasmas in the presence of strong magnetic
  fields},'' \href{http://dx.doi.org/10.1103/PhysRevD.94.054020,
  10.1103/PhysRevD.96.019903}{{\em Phys. Rev.} {\bfseries D94} no.~5, (2016)
  054020}, \href{http://arxiv.org/abs/1605.06061}{{\ttfamily arXiv:1605.06061
  [hep-ph]}}.
[Erratum: Phys. Rev.D96,no.1,019903(2017)].

\bibitem{Hattori:2017usa}
K.~Hattori, Y.~Hirono, H.-U. Yee, and Y.~Yin, ``{MagnetoHydrodynamics with
  chiral anomaly: phases of collective excitations and instabilities},''
  \href{http://dx.doi.org/10.1103/PhysRevD.100.065023}{{\em Phys. Rev.}
  {\bfseries D100} no.~6, (2019) 065023},
\href{http://arxiv.org/abs/1711.08450}{{\ttfamily arXiv:1711.08450 [hep-th]}}.

\bibitem{Grozdanov:2017kyl}
S.~Grozdanov and N.~Poovuttikul, ``{Generalised global symmetries in
  holography: magnetohydrodynamic waves in a strongly interacting plasma},''
  \href{http://dx.doi.org/10.1007/JHEP04(2019)141}{{\em JHEP} {\bfseries 04}
  (2019) 141},
\href{http://arxiv.org/abs/1707.04182}{{\ttfamily arXiv:1707.04182 [hep-th]}}.

\bibitem{Hofman:2017vwr}
D.~M. Hofman and N.~Iqbal, ``{Generalized global symmetries and holography},''
  \href{http://dx.doi.org/10.21468/SciPostPhys.4.1.005}{{\em SciPost Phys.}
  {\bfseries 4} no.~1, (2018) 005},
\href{http://arxiv.org/abs/1707.08577}{{\ttfamily arXiv:1707.08577 [hep-th]}}.

\bibitem{Jensen:2012kj}
K.~Jensen, R.~Loganayagam, and A.~Yarom, ``{Thermodynamics, gravitational
  anomalies and cones},'' \href{http://dx.doi.org/10.1007/JHEP02(2013)088}{{\em
  JHEP} {\bfseries 02} (2013) 088},
\href{http://arxiv.org/abs/1207.5824}{{\ttfamily arXiv:1207.5824 [hep-th]}}.

\bibitem{1202.2161}
G.~Basar and D.~E. Kharzeev, ``{The Chern-Simons diffusion rate in strongly
  coupled N=4 SYM plasma in an external magnetic field},''
  \href{http://dx.doi.org/10.1103/PhysRevD.85.086012}{{\em Phys. Rev.}
  {\bfseries D85} (2012) 086012},
\href{http://arxiv.org/abs/1202.2161}{{\ttfamily arXiv:1202.2161 [hep-th]}}.

\end{thebibliography}\endgroup
\end{document}